\documentclass[11pt,a4paper]{article}
\DeclareMathAlphabet{\scr}{U}{rsfs}{m}{n}

\pdfoutput=1
\usepackage[utf8]{inputenc}
\usepackage{latexsym}
\usepackage{epsfig}
\usepackage[mathscr]{eucal}
\usepackage{amsfonts}
\usepackage{amscd}
\usepackage{cite}
\usepackage{array}   
\usepackage{amssymb}
\usepackage{colordvi}
\usepackage[centertags]{amsmath}
\usepackage{enumerate}
\usepackage{graphicx}
\usepackage{booktabs}
\usepackage{hyperref}
\usepackage{enumitem}
\setlist[description]{leftmargin=2\parindent,labelindent=\parindent}
\usepackage{theorem}
\usepackage{listings}
\usepackage{multirow}
\usepackage[framemethod=TikZ]{mdframed}
\usepackage{makecell}
\usepackage[footnotesize]{caption}
\usepackage{bbm}
\usepackage[labelfont=bf,textfont={sl,bf}]{subfig}
 \usepackage{slashed}
\usepackage{xcolor}
\xdefinecolor{mygray}{gray}{0.85}
\xdefinecolor{mygray}{gray}{0.85}
\usepackage{bm}
\usepackage{arydshln}
\usepackage{pifont}
\usepackage{mathtools}
\usepackage{siunitx}
\mdfsetup{%
   middlelinecolor=black,
   middlelinewidth=0.5pt,
   roundcorner=0pt,
   innertopmargin=-3pt,
   innerbottommargin=10pt
}

\newcommand{\newc}{\newcommand}
\newc{\beq}{\begin{eqnarray}}
\newc{\eeq}{\end{eqnarray}}
\newc{\ol}{\overline}
\newc{\bs}{\boldsymbol}
\newc{\m}{\mathcal}
\newc{\lan}{\langle}
\newc{\ra}{\rangle}
\newc{\pa}{\partial}
\newc{\nn}{\nonumber}

\newcommand{\hs}{\hat{s}}
\newcommand{\hu}{\hat{u}}
\newcommand{\htt}{\hat{t}}

\newcommand{\green}[1]{{\color[rgb]{0.1, 0.5, 0.1} #1}}

\begin{document}
\title{
\vspace*{-3cm}
\phantom{h} \hfill\mbox{\small KA-TP-20-2023} \\[-0.25cm]
\phantom{h} \hfill\mbox{\small TU-1212}
\\[1cm]
\textbf{Composite 2-Higgs Doublet Model:\\[0.25cm]
Strong Effects on Higgs Pair Production}}

\date{}
\author{Stefania De Curtis$^{1\,}$\footnote{E-mail:
  \texttt{decurtis@fi.infn.it}}, ~Luigi Delle Rose$^{2\,}$\footnote{{E-mail:
  \texttt{luigi.dellerose@unical.it}}}, ~Felix Egle$^{3\,}$\footnote{E-mail:
  \texttt{felix.egle@kit.edu}}, \\[0.25cm] Stefano Moretti$^{4,5\,}$\footnote{E-mail:
  \texttt{s.moretti@soton.ac.uk; stefano.moretti@physics.uu.se}}, ~Margarete M\"{u}hlleitner$^{3\,}$\footnote{E-mail:
  \texttt{margarete.muehlleitner@kit.edu}},
~Kodai Sakurai$^{6,7\,}$\footnote{E-mail: \texttt{kodai.sakurai@fuw.edu.pl}}
\\[5mm]
{\small \it  $^1$INFN Sezione di Firenze and GGI,}\\
{\small\it Department of Physics and Astronomy, University of Florence} \\
{\small \it  Via Sansone, 1, 50029 Sesto Fiorentino (FI), Italy}\\[3mm]
{\small \it $^2$Dipartimento di Fisica, Universita’ della Calabria,}\\
{\small\it I-87036 Arcavacata di Rende, Cosenza, Italy}\\
{\small\it INFN-Cosenza, I-87036 Arcavacata di Rende, Cosenza, Italy}\\[3mm]
{\small \it $^3$Institute for Theoretical Physics, Karlsruhe Institute of Technology,} \\
{\small \it Wolfgang-Gaede-Str. 1, 76131 Karlsruhe, Germany}\\[3mm]
{\small \it $^4$School of Physics and Astronomy, University of Southampton,} \\
{\small \it Highfield, Southampton SO17 1BJ, UK}\\[3mm]
{\small \it $^5$Department of Physics and Astronomy, Uppsala University,} \\
{\small \it Box 516, SE-751 20 Uppsala, Sweden}\\[3mm]
{\small \it $^6$Institute of Theoretical Physics, Faculty of Physics, University of Warsaw, } \\
{\small \it ul.~Pasteura 5, PL-02-093 Warsaw, Poland} \\[3mm]
{\small \it $^7$Department of Physics, Tohoku University, } \\
{\small \it Sendai, Miyagi 980-8578, Japan}}
\maketitle
\vspace*{-0.85truecm}
\begin{abstract}
{
\noindent \small
We show how effects of compositeness emerging in a Composite 2-Higgs Doublet Model can enter Standard Model (SM)-like 
Higgs boson pair production at the Large Hadron Collider in both resonant and non-resonant mode. 
Such effects can arise from modified trilinear Higgs self-couplings and top-Yukawa couplings, as well as from loops of new heavy quarks and additional quartic Higgs-fermion interactions. In the resonant case, significant distortions of the Breit-Wigner shape of a new scalar state decaying into the two SM-like Higgs states
 may occur due to interference effects amongst not only the SM-like diagrams but also those involving the new heavy quarks. In the non-resonant case, a modification of the underlying line-shape and a local maximum at twice a new heavy quark mass appear simultaneously. We quantify these effects by taking into account the relevant theoretical and latest experimental bounds.} 
\end{abstract}
\thispagestyle{empty}
\vfill
\newpage

\section{Introduction}
\label{sec:Introduction}
During run 1 and 2 of the Large Hadron Collider (LHC), precise measurements of the discovered Higgs boson properties, namely, its quantum numbers and couplings to Standard Model (SM) particles, have revealed these to  be very SM-like \cite{ATLAS:2022vkf,CMS:2022dwd}. However,
in order to ultimately establish that the Brout-Englert-Higgs mechanism
\cite{Higgs:1964pj,Englert:1964et,Guralnik:1964eu,Kibble:1967sv} is
indeed responsible for the generation of elementary particle masses, one really needs to reconstruct the Higgs
potential itself by measuring the Higgs self-couplings~\cite{Djouadi:1999gv,Djouadi:1999rca}.
With increasing amounts of data, the two multi-purpose LHC experiments, ATLAS and CMS, have started to put limits on the trilinear Higgs self-coupling by investigating both resonant and non-resonant SM-like Higgs pair production. In the absence of a discovery, the measured limits are rather  weak, as they constrain the trilinear Higgs self-coupling to values between $-0.4$
and $6.3$ times the SM value as reported by ATLAS
\cite{atlaspaperdihiggs} and between $-1.7$ and $8.7$ times the SM value
as found by CMS \cite{CMS:2022hgz}  (both assuming a SM-like
top-Yukawa coupling). 

The reason for the limited LHC sensitivity to SM-like Higgs pair production, compared to  single Higgs production, is the smallness of the corresponding cross section. At the LHC, the dominant production process is induced by gluon-gluon fusion also in the case of pair production \cite{Baglio:2012np,deFlorian:2016spz,DiMicco:2019ngk}. The Leading-Order (LO) amplitude in the SM is built up by triangle and box diagrams mediated by heavy quark loops \cite{Glover:1987nx,Dicus:1987ic,Plehn:1996wb}. Not only loop-mediated, but also suffering from a destructive interference between box and triangle loops,  this implies 
a small SM cross section of $\sim 31$~fb for a center-of-mass energy
of $\sqrt{s}=13$~TeV  \cite{Grazzini:2018bsd}\footnote{This number corresponds to the FT$_{\text{approx}}$ value, the current state-of-the-art in perturbative QCD, wherein the cross section is computed at Next-to-Next-to-LO (NNLO)  in the heavy-top limit with full LO and NLO mass effects and full mass dependence in the one-loop double real corrections at NNLO.}.

While it may not be possible to measure the trilinear Higgs self-coupling in the SM during run 3, for which the High-Luminosity LHC (HL-LHC) \green{\cite{Gianotti:2002xx,Cepeda:2019klc}} will then be required, in Beyond-the-SM (BSM) scenarios with extended Higgs sectors this may well be achievable already in a few years from now. Here, the cross sections can be enhanced due to new heavy (resonant) Higgs states,
additional particles running in the loop, modified trilinear Higgs self-couplings and/or top-Yukawa couplings (see, e.g., \cite{Abouabid:2021yvw} for a recent comprehensive study of various archetypal BSM extensions). Taking into account all relevant theoretical and  experimental constraints, the trilinear Higgs self-coupling of the SM-like Higgs boson in BSM extensions can still deviate significantly from the SM value {while} the top-Yukawa coupling is constrained to be within about $\pm 10$\% of  the SM value~\cite{Abouabid:2021yvw}. 

One of the simplest Higgs sector extensions that may allow for SM-like Higgs pair production embedding all aforementioned effects
is given by the 2-Higgs Doublet Model (2HDM) \cite{Lee:1973iz,Branco:2011iw}\footnote{For the extension of the Effective Field Theory (EFT) framework to the 2HDM, see \cite{Crivellin:2016ihg}.}, where a second Higgs doublet is added to the SM Higgs sector. While the elementary version of the 2HDM can naturally be accommodated in supersymmetry in the form of the Minimal Supersymmetric Standard Model (MSSM) (see, e.g.,~\cite{Gunion:1989we,Martin:1997ns,Dawson:1997tz,Djouadi:2005gj,Moretti:2019ulc} for reviews), when its Yukawa couplings are of so-called Type-II, it was recently proposed that a counterpart version of this Higgs scenario realised within compositeness, the so-called Composite 2HDM (C2HDM) of \cite{DeCurtis:2018zvh}, wherein the Yukawa couplings are aligned in flavour space, 
could yield a very distinctive phenomenology in comparison, so that one may well be able to tell the two apart \cite{DeCurtis:2018iqd}. 

In the spirit of testing compositeness further, it is thus of clear importance to study SM-like Higgs pair (di-Higgs, for short) production within it on the same footing as done in \cite{Moretti:2023dlx} in the context of supersymmetry. While the MSSM and C2HDM offer the same Higgs sector in terms of particle content, they differ in the mechanism that enables one to overcome the hierarchy problem of the SM. In order to explain the relatively small SM-like Higgs mass value in the presence of large mass scales, supersymmetry invokes spin-0 companions to the top-quark, the top squarks, with the latter adequately cancelling the otherwise excessive corrections to the SM-like Higgs mass due to the former.  Compositeness resorts instead to an alternative mechanism, wherein a SM-like Higgs state is naturally light since it is conceived as a 
pseudo-Nambu-Goldstone Boson (pNGB) with composite nature generated in a new strong sector, {which} dynamics also entails the presence (among others) of new spin-1/2 states,  i.e., top-quark companions. These states, too, can then enter di-Higgs production via the aforementioned loop diagrams\footnote{Effects of dark-coloured scalar particles on di-Higgs production, assuming a SM-like Higgs sector, were recently investigated in \cite{Gabriel:2023dyx}.}. It is the purpose of the present study to assess the role of such extended Higgs and fermionic sectors of the C2HDM in di-Higgs production at the LHC (for earlier work along the lines of our paper, see e.g. \cite{Cheung_2021}). In doing so, we will treat both cases of resonant and non-resonant di-Higgs production (separately), following a high-intensity scan of the C2HDM parameter space in presence of all available theoretical and experimental constraints, which will enable us to define viable Benchmark Points (BPs), with different dynamical and/or kinematical characteristics,  amenable to further phenomenological investigation. In fact, while the results of the present analysis are primarily of theoretical nature, they serve to demonstrate that a computable framework exists within compositeness that can eventually be tested experimentally, thereby ultimately contributing to put the C2DHM on a similar footing as the MSSM, while also being notably different from elementary 2HDM realisations.

The outline of the paper is as follows. In the next section we will introduce the C2HDM together with the constraints that we applied on it. In Sec.~\ref{sec:dihiggcxn}, we will give the di-Higgs cross section for SM-like Higgs pair production in such a scenario. Subsequently, in Sec.~\ref{sec:numerical}, we will present our numerical results: we will first describe our parameter scan with the applied constraints and  then move on to discuss the origin and impact of C2HDM effects. In Secs.~\ref{sec:nonresonant} and \ref{sec:resonant}, we will then investigate in detail the non-resonant and resonant Higgs pair production case, respectively, both for inclusive and exclusive di-Higgs production. Sec.~\ref{sec:comparison} discusses the differences between results in the C2HDM, the elementary 2HDM Type-II, the elementary flavour-Aligned 2HDM (A2HDM) and the MSSM. Our conclusions are given in Sec.~\ref{sec:summary}.

\section{Explicit Realisation of the C2HDM}

\label{sec:Model}

As indicated above, compositeness can naturally solve the hierarchy problem of the SM 
by the introduction of a light Higgs boson emerging as a pNGB from a strongly-coupled sector (for a review, see, e.g.,~\cite{Contino:2010rs}). It is a bound state produced by strong dynamics \cite{Dimopoulos:1981xc,Kaplan:1983sm,Kaplan:1983fs,Banks:1984gj,Georgi:1984ef,Georgi:1984af,Dugan:1984hq} and, due to its Goldstone nature, it is separated by a mass gap from the other usual  resonances of the strong sector. It can be seen as the analogue of the pion of the strong interactions and, 
just like QCD predicts other mesons as bound states, there could be several Higgs-like states predicted by compositeness beyond the one discovered so far as well as additional composite states, both fermionic and bosonic ones. In this respect, a natural setting is the C2HDM of Refs.~\cite{Mrazek:2011iu, DeCurtis:2016scv, DeCurtis:2016tsm,DeCurtis:2017gzi, DeCurtis:2018zvh}. It is built by enlarging the coset associated to the breaking of the global symmetry of the underlying strong interactions to contain, besides the SM-like Higgs doublet, an additional one (both realised through pNGB states). The presence of an extra Higgs doublet is also predicted in supersymmetric models. A comparative study between the C2HDM and  MSSM has been presented in \cite{DeCurtis:2018iqd} for the case of single SM-like Higgs boson  production. In the following phenomenological analysis we will tension the C2HDM to its supersymmetric counterpart, i.e., the MSSM
(albeit limitedly  to the non-resonant case),  
by referring the reader to \cite{Moretti:2023dlx}, as well as two elementary 2HDMs, the so-called Type-II (with a Higgs sector similar to that of the MSSM, yet, with more degrees of freedom) and the A2HDM (which realises a Yukawa structure similar to that of the C2HDM).

Here, we focus on the extended Higgs sector of the C2HDM, which is  originating from the breaking $ SO(6)\to SO(4) \times SO(2)$   at the compositeness scale $f$ providing two pNGB doublets (see \cite{DeCurtis:2018zvh} for the explicit construction of the model).
The pNGB matrix $U$ is constructed from the 8 broken $SO(6)$ generators $T^{\hat{a}}_{i}$ ($i=1,2$, $\hat{a}=1,\dots,4$) out of the 15 total ones $T^A$ ($A=1,\dots, 15$) as follows:
\
\begin{align}
&U=e^{i\frac{\Pi}{f}},\quad
\Pi\equiv \sqrt{2}\phi_{i}^{\hat{a}}T^{\hat{a}}_{i}=-i 
\begin{pmatrix}
0_{4\times 4} & {\bf \Phi} \\
-{\bf \Phi}^T & 0_{2\times 2}  
\end{pmatrix},   \label{u6}
\end{align}
where ${\bf \Phi} \equiv (\phi_1^{\hat{a}},\phi_2^{\hat{a}}) $. 
The two real 4--vectors $\phi_i^{\hat{a}}$ can be rearranged into two complex doublets $\Phi_i$ ($i=1,2$) as 
\begin{align}
\Phi_i = \frac{1}{\sqrt{2}}\left(
\begin{array}{cc}
\phi_i^{\hat 2} + i\phi_i^{\hat 1} \\
\phi_i^{\hat 4} - i\phi_i^{\hat 3} 
\end{array}\right). \label{doublet6}
\end{align}
The Vacuum Expectation Values (VEVs) of the Higgs fields 
are taken to be $\langle \phi_i ^{\hat 4}\rangle = v_i$ ($i=1,2$),  with $v^2 = v_1^2 + v_2^2$ and the definition $\tan\beta \equiv v_2/v_1$ as usual in  2HDMs, such that $\phi_i ^{\hat 4} = v_i + h_i$.
Because of the non-linear nature of the pNGBs eventually emerging as Higgs boson states, $v$ is related  to the SM Higgs VEV $v_{\text{\rm SM}}^{}$ via  
\begin{align}
v_{\text{\rm SM}}^2 = f^2\sin^2 \frac{v}{f} 
\end{align}
and it reduces to it in the $f\to\infty$ limit.

To obtain a non-zero value for the Higgs boson masses, terms which break the $SO(6)$ symmetry explicitly must be introduced.  
Within the partial compositeness paradigm~\cite{Kaplan:1991dc},  this can be achieved by a linear mixing between the (elementary) SM  and (composite)  strong sector fields. In our construction, this is implemented for  the third generation of the SM fermions, which are in fact the main sources of the new symmetry breaking.  The masses of the Higgs bosons and their self-interactions are then generated at one-loop level from the Coleman-Weinberg potential. As such, {these}  are not free parameters, unlike in  elementary realisations of a 2HDM, but depend upon the strong sector dynamics and present  correlations amongst themselves. 

The explicit model is  based on the two-site construction defined in \cite{DeCurtis:2011yx}.
The extra degrees of freedom in the gauge  sector are the spin-1 resonances of the adjoint of $SO(6)$, that we will consider very heavy and so  phenomenologically irrelevant for the purposes of the present analysis. Their Lagrangian ${\cal L}_\text{gauge}$ is explicitly given in Ref.~\cite{DeCurtis:2018zvh} and depends on the gauge coupling constant $g_\rho$ of the strong sector. This term is not relevant here and will not be detailed below.
In contrast, in the fermion sector, we consider two spin-$1/2$ resonances $\Psi^{I/J}$ ($I,J=1,2$) which are $SO(6)$ ${\bf 6}$-plets, which can be much lighter in comparison. This is the minimum  number of fermionic multiplets required to ensure the UV finiteness of the potential \cite{DeCurtis:2018zvh}.
The corresponding Lagrangian is 
\begin{eqnarray}
{\cal L}_{\text{strong}}^{\text{ferm}}  + {\cal L}_{\text{mix}}^{\text{ferm}}  &=& 
 \bar{\Psi}^I iD\hspace{-2.2mm}/\hspace{0.6mm} \Psi^I +[
 - \bar{\Psi}_{L}^I M_\Psi^{IJ} \Psi_R^J  - \bar{\Psi}_L^I (Y_1^{IJ} \Sigma + {Y}_2^{IJ} \Sigma^2 )\Psi_R^J \notag\\ &&  + (\Delta_L^{I}\bar{q}_L^{\bf{6}} \Psi_R^I +  \Delta_R^I \bar{t}_R^{\bf{6}} \Psi_L^I )]+ 
\text{h.c.}, \label{ferm}
\end{eqnarray}
where $\Sigma = U \Sigma_0U^T$, with  
 $\Sigma_0 = \text{diag}(0_{4\times 4}, i\sigma_2)$. 
  Here, $\Delta_{L,R}$ ($M_\Psi$, $Y_1$ and ${Y}_2$) are dimensionful  vectors ($2\times 2$ matrices).
Since $\Sigma^3 = -\Sigma$, terms up to a quadratic power in $\Sigma$ reproduce the most general interaction Lagrangian between the fermionic resonances and the pNGB fields. For simplicity, we assume CP conservation in the strong sector, i.e., all parameters in Eq.~(\ref{ferm}) are  real (the CP-violating case is considered in \cite{DeCurtis:2021uqx}).
As a result, the  Higgs potential is CP-conserving. The Yukawa interactions 
for the right-handed bottom quark can also be included by introducing further spin-1/2 resonances.
Analogously, one can implement  partial compositeness for $\tau$ leptons, too. {In the following analysis of di-Higgs production, though, we will only consider the top sector as being composite amongst fermions, with the relevant states given by the ordinary SM top quark and eight heavy top partners of electric charge 2/3.}
We finally note that the new gauge interactions do not give rise to severe  
UV divergences in the calculation of the Higgs potential. In fact,  the fermion Lagrangian defined above provides logarithmic UV divergences which can be removed by suitable conditions amongst the strong sector parameters. 

 In Ref.~\cite{DeCurtis:2018zvh}, among all possible solutions, we  enforced  the ``left-right symmetry'' defined in \cite{DeCurtis:2011yx}, which provides a simple setup, but represents only a part of the available parameter space of the C2HDM. Thus, here, 
 we enlarge the C2HDM parameter space by removing the relations among the new fermion parameters ensuing from such a symmetry. Namely, we  consider 
$\Delta_L^{I}, \Delta_R^{I}, Y_{1,2}^{IJ}$ and $M_\Psi^{IJ}$
as parameters of the fermion sector in their full flavour structure, naturally of the order of  the compositeness scale $f$, and we require the UV finiteness of the Higgs potential at one-loop order by enforcing the vanishing of the coefficients of the quadratic and logarithmic divergences. 
To satisfy these conditions, as said above, at least two families of heavy fermions are needed, and we opted for the minimal realisation with $I,J=1,2$. 
Moreover, the finiteness equations are given as combinations of the parameters listed above and allow to effectively reduce the number of independent ones. For example, one can compute the mass parameter $M_\Psi^{IJ}$ in terms of all the others, thus leaving the fermion sector completely determined by 
\begin{equation}
 \Delta_L^{I}, \quad \Delta_R^{I} , \quad Y_{1,2}^{IJ} \quad (I,J=1,2)\\
\label{UV2}
\end{equation}
\noindent 
as free parameters.

By integrating out the new spin-1  and -1/2 states, 
we obtain the effective low-energy Lagrangian for the SM (gauge) bosons and (third generation) fermions as well as the Higgs fields $\Phi_i$ ($i=1,2$). 
In each coefficient of the Lagrangian terms,  form factors then  appear, which are expressed as functions of the parameters of the strong sector. 
Their explicit forms are provided in \cite{DeCurtis:2018zvh}. However,
the low energy Lagrangian  introduces, in general, Flavour-Changing Neutral Currents (FCNCs) at tree level via Higgs boson exchanges. Here, in order to avoid these FCNCs, we will follow \cite{DeCurtis:2018zvh} by requiring alignment in the flavour sector, like  in the elementary A2HDM. 

The Higgs potential is then generated  by the gauge boson ($W_\mu^a$, $a=1,2,3$, and $B_\mu$) 
and  fermion ($q_L$ and $t_R$) loop contributions.  
By expanding  up to the fourth order in the  $\Phi_i$  fields, we get
a Higgs potential with the same structure as the one of  the elementary A2HDM but with coefficients
determined in terms of the parameters of the strong sector. 
Therefore, the masses of the Higgs bosons and the scalar mixing angle are fully predicted by the strong dynamics.
Phenomenologically acceptable configurations are obtained by requiring the vanishing of the two
tree-level tadpoles as well as the reproduction of the Electro-Weak (EW) VEV and the measured masses of the top quark and the SM-like Higgs boson, here identified as the lightest CP-even scalar state.

\subsection{Constraints on the C2HDM \label{sec:constraints}}

The composite nature of the SM-like Higgs boson in the C2HDM can be accessed at the LHC by exploiting the corrections of ${\cal O} (\xi)$  to its couplings, where $\xi= v^2_{\rm SM}/f^2$.  The  precision achieved by the current LHC data is at the 10--20\% level depending on the involved  interactions   
\cite{ATLAS:2022vkf,CMS:2022dwd}.
Specifically, the ${\cal O} (\xi)$ corrections to the Yukawa interactions with top and bottom quarks or $\tau$ leptons are notoriously difficult to measure at the LHC as they are affected by a significant QCD background (further, leptonic $\tau$ decays are subleading). A much cleaner alternative is to probe the interactions of the SM-like Higgs with the gauge bosons, which are also affected by similar corrections. 
The model is thus tested against experimental measurements through the latest version of the \texttt{HiggsBounds} \cite{Bechtle:2013wla} and \texttt{HiggsSignals} \cite{Bechtle:2013xfa} packages, 
which test the null results of extra Higgs boson searches and compatibility with the Higgs data, respectively. Hereafter, in all BSM scenarios that we will consider, the SM-like Higgs boson will always be provided by the lighter CP-even Higgs state $h$ (recall that, conventionally, $m_h<m_H$, with $H$ denoting the second CP-even Higgs mass eigenstate).

\begin{figure}
    \centering
    \includegraphics[width=0.45\textwidth]{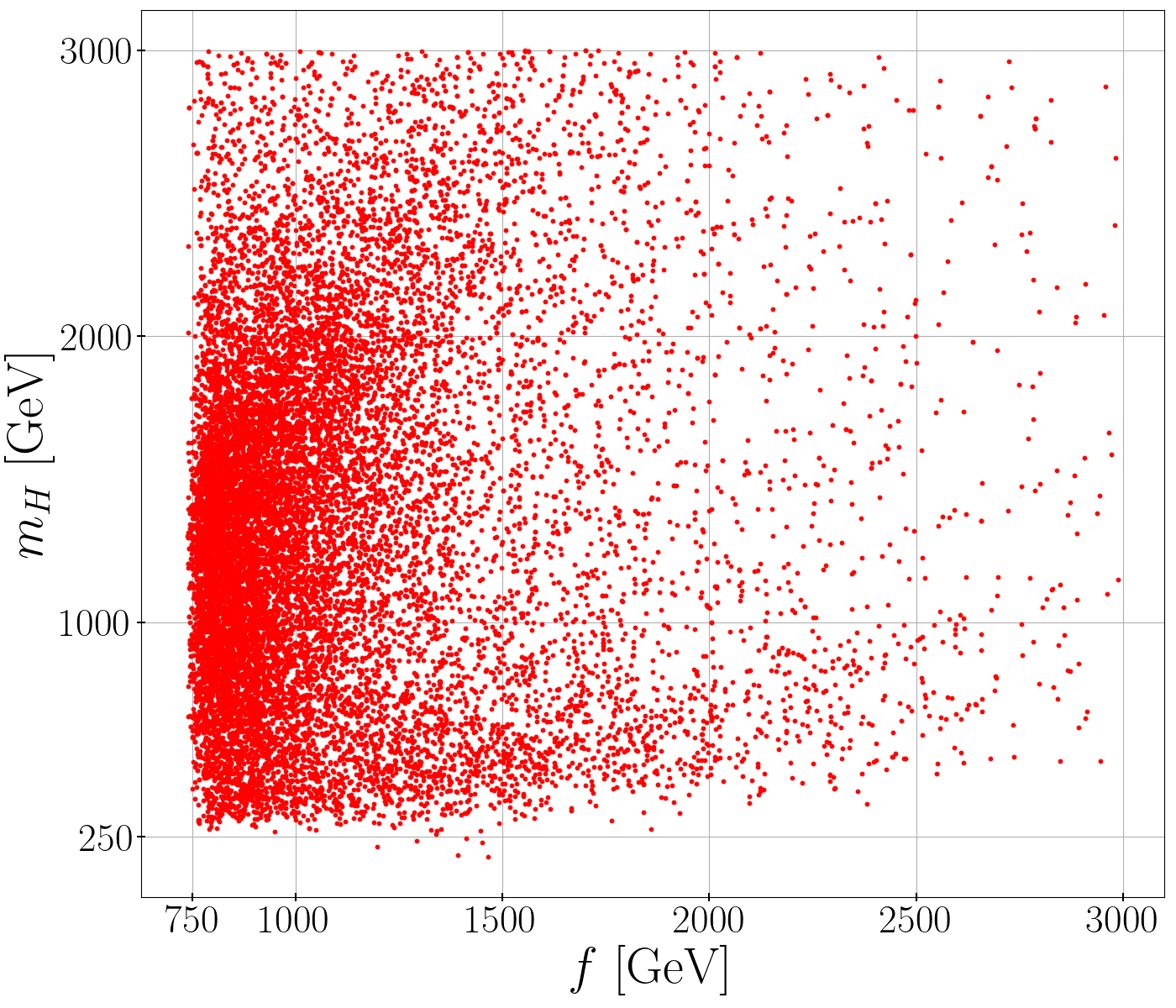}
    \caption{Predicted values for $m_H$ as function of the compositeness scale $f$ in the C2HDM.  The points satisfy current experimental constraints. }
    \label{fmH}
\end{figure}

In Fig.~\ref{fmH}, we show the predicted values for $m_H$ as function of the compositeness scale $f$. The red points satisfy the  constraints from the current direct and indirect Higgs searches. The lower bound of $f \approx 750$ GeV derives mainly from the Higgs coupling measurements.
We can extrapolate from the present $\kappa$ values extracted from the LHC data of about 30 fb$^{-1}$  of integrated luminosity  to 300 fb$^{-1}$ (end of Run 3) and 3000 fb$^{-1}$ 
 (HL-LHC) by adopting the expected experimental accuracy  (as done in \cite{DeCurtis:2019jwg}). In particular, an estimate of the extrapolated lower bound on $f$ can simply be obtained by the  expected precision on the Higgs coupling measurements which will be reached after the HL-LHC. Following Ref.~\cite{deBlas:2019rxi}, we expect an improvement to a precision of 2--4\%, which translates to $f\ge 1.2$ TeV. This means that, on the one hand,  CHMs will not be ruled out anytime soon by  precision measurements of the Higgs couplings and, on the other hand, that there will be the chance to observe new signals or deviations from other precision measurements due to our framework before the end of the HL-LHC phase. In the end, we decided to scan $f$ up to 3 TeV because, as it is clear from the plot, the number of points decreases for higher values of the compositeness scale since a large fine-tuning is needed to recover the top quark and SM-like Higgs mass values.
 
As stated above, in the C2HDM, the entire scalar potential is generated by the strong dynamics and, as such, all mass terms $m_i^2$ and quartic couplings $\lambda_i$ are determined by the parameters of the strong sector. We required the potential to reproduce the correct vacuum structure with a global EW minimum, as well as to be compliant with perturbativity constraints, namely $\lambda_i < \sqrt{4 \pi}$. Concerning the constraints from the EW precision measurements we refer to \cite{DeCurtis:2018zvh}. 

The C2HDM predicts the presence of new resonances as composite states from the strong dynamics. We will not consider here the heavy spin-1 resonances: their mass can naturally be  in the multi-TeV region so that they can safely be integrated out. On the contrary, the top quark partners play a crucial role in the di-Higgs production cross section. 
Bounds on their masses are derived from their pair production searches at the LHC. In fact, values extracted from experimental data of their  production cross sections $\sigma$ times Branching Ratio (BR) set limits depending on the extra-fermion $T_{2/3}$ mass and any  assumption on its BRs. {The experimental analyses assume, however, that the $T_{2/3}$ solely  decays into SM particles \cite{Benbrik:2019zdp}.} 
This is not the case for the extra fermions entering  the C2HDM. In fact, a $T_{2/3}$ state can decay also in exotic channels, like $Ht$, $At$ and $H^+ \bar b$, 
{where $A$ and $H^+$ denote the CP-odd and  charged Higgs boson of the C2HDM Higgs sector, respectively.} They can have sizeable BRs, thus potentially relaxing the experimental bounds \cite{Benbrik:2019zdp}. Yet, following the recent analysis of Ref.  \cite{ATLAS:2022tla}, it seems unlikely that the mass of the {lightest among the} heavy top quark {partners} can be less than 1.3 TeV or so. To be on the safe side, we will thus consider only BPs with $m_{Ti}\ge$ 1.3 TeV,   with $i=1,...,8$  labelling the eight heavy top quark states which are active in the loop contributions to the $gg\to hh$ process, as described in the following. 

Modifications to rare flavour transitions in the SM are potentially present in our model due to exchange of the pNGBs. Indeed, the heavy Higgses can mediate tree-level corrections in charged current processes and loop corrections in the neutral ones. However, since the scalars have flavour aligned interactions, all the flavour constraints are due to a rescaling of the corresponding SM rates.
In \cite{DeCurtis:2018zvh} we have identified the most relevant flavour processes that may constrain our model. The bounds from charged meson decays are usually important only for small charged Higgs masses or for large couplings and, as such, the corresponding excluded region does not overlap with the parameter space considered here. Similarly, the constraints from the $B_s \to \mu^+ \mu^-$ transition do not affect the region of masses and couplings relevant for our analysis. The most relevant constraints may in fact come from the $b \to s \gamma$ transition, which depends on the interplay between the top and bottom couplings of the charged Higgs boson, which are determined by the degree of $C_2$ breaking\footnote{The discrete $C_2$ transformation is realised on the Higgs sector as $H_1 \to H_1$ and $H_2 \to -H_2$ and is akin to the $Z_2$ one in the elementary case.}, $\zeta_{b,t}$, in each sector. As shown in \cite{DeCurtis:2018zvh}, the bound can  easily be relaxed by requiring $\zeta_b \lesssim 0.1 \zeta_t$.


\section{Higgs Pair Production in the C2HDM \label{sec:dihiggcxn}}

\subsection{Building Blocks of di-Higgs Production in the C2HDM}
At the LHC, the  dominant Higgs pair production process is given by gluon fusion, which is mediated by heavy quark loops. Due to partial compositeness, entailing further heavy quarks beyond the SM ones, we have to specify the heavy quark sector of our model. The fermion Lagrangian is given in Eq.~(\ref{ferm}) (see \cite{DeCurtis:2018zvh} for more details).
As already stated, in order to ensure the finiteness of the potential, we need at least two families of heavy fermions. For our analysis, we opt for the minimal setup with $\Psi^I$ ($I = 1, 2$). The ensuing fermion sextuplets are then decomposed as\footnote{Hereafter, we use the short-hand notations $s_\theta\equiv \sin\theta$ and $c_\theta\equiv\cos\theta$.}
\begin{align}
\label{eq:multi_dec}
q_L^{\bm 6} = \frac{1}{\sqrt{2}} \left( \begin{array}{c} i b_L \\ b_L \\ i t_L \\ - t_L \\ 0 \\ 0  \end{array} \right) \,, \qquad
t_R^{\bm 6} =  \left( \begin{array}{c} 0 \\ 0  \\ 0   \\ 0 \\ c_{\theta_t} \\ i  s_{\theta_t}  \end{array} \right) t_R   \,, \qquad
\Psi = \left( \begin{array}{c} \psi_4 \\ \psi_2 \end{array} \right) = \frac{1}{\sqrt{2}} \left( \begin{array}{c}
i B_{-1/3} - i X_{5/3} \\  B_{-1/3} +  X_{5/3}  \\   i T_{2/3} + i X_{2/3}  \\  - T_{2/3} +  X_{2/3}  \\ \sqrt{2}  \tilde T_1  \\ \sqrt{2}  \tilde T_2
 \end{array} \right),
\end{align}
with the angle $\theta_t$ controlling the embedding of the right-handed top quark in the fundamental representation of $SO(6)$. Here, we choose $\theta_t = 0$, which ensures a CP-invariant realisation of the Higgs potential. The multiplet decomposition of $\Psi$ in Eq.~(\ref{eq:multi_dec}) holds for both the extra sextuplets.
Therefore, we have as additional fermionic content:
\begin{itemize}
\item 4 top partners with $Q = 2/3$:  $X_{2/3}$, $T_{2/3}$, $\tilde T_1, \tilde T_2$;
\item 1 bottom partner with $Q= -1/3$: $B_{-1/3}$;
\item 1 exotic fermion with $Q = 5/3$: $X_{5/3}$.
\end{itemize}

The count above must be doubled due to the presence of two families of heavy fermions ($\Psi^I$, $I=1,2$).
While the diagonalisation of the mass matrices of the bottom partners and of the exotic fermions is trivial, the corresponding one for the 9  fermions with charge 2/3 (i.e., the 8 top quark partners plus the ordinary SM top quark) can only be done numerically.

\begin{figure}[tb]
\centering
\includegraphics[width=\textwidth]{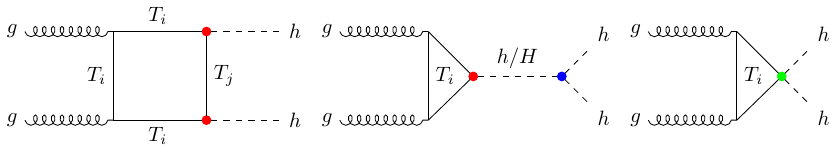}
\vspace*{-0.5cm}
\caption{Generic diagrams contributing to SM-like $hh$ production via gluon fusion in the C2HDM, mediated by the SM top (denoted by $T_9$) and the heavy top partner (denoted by $T_1,...,T_8$) loops {($i,j=1,...,9$)}. The coloured dots indicate the new and/or modified interactions w.r.t.~the SM.} 
\label{fig:LODiagrams}
\end{figure}

In Fig.~\ref{fig:LODiagrams}, the generic Feynman diagrams contributing to SM-like Higgs pair production via gluon fusion in the C2HDM are depicted. In the box and in the triangle diagrams all 9 fermions with $Q=2/3$ enter. 
The first two diagrams  involve the modified (by corrections of order $\xi$) SM-like Higgs-top Yukawa coupling and the new $h$ couplings to the heavy top partners (denoted by the red dots).  The second diagram takes into account the  exchange of the extra Higgs state $H$, involving the new trilinear coupling $\lambda_{Hhh}$. Also, the trilinear coupling of the SM-like Higgs, $\lambda_{hhh}$, which appears for the $h$ exchange in the $s$-channel, is modified w.r.t. the SM (by corrections of order $\xi$). The new, respectively modified, trilinear Higgs self-couplings are marked by the blue dot. Finally, in the C2HDM, we have novel quartic couplings (denoted by the green dot) between two Higgs bosons and two fermions, leading to the third diagram. Such quartic couplings are typical of theories with Higgs states as pNGBs.

For the derivation of the Feynman rules, we give here the relevant Lagrangians. The Lagrangian involving the needed scalar self-interactions reads
\begin{align}
\mathcal L_\textrm{scalar}^\textrm{int} = & - \frac{1}{3!} \lambda_{hhh} h^3   - \frac{1}{2} \lambda_{hhH}^{(1)} h^2 H \nn\\ 
+& \frac{v}{3 f^2} (h_2 \partial_\mu h_1 - h_1 \partial_\mu h_2) \partial^\mu h_2 
+ \cdots, \label{eq:lscalarint}
\end{align}
where 
\begin{align}
    \left( \begin{array}{c} h_1 \\ h_2 \end{array} \right) = 
\left( \begin{array}{cc} c_\theta & - s_\theta \\ s_\theta & c_\theta \end{array} \right) \left( \begin{array}{c} h \\ H \end{array} \right)  \,.
\end{align}
The terms in  the last line of Eq.~(\ref{eq:lscalarint}), which are typical of  Composite Higgs Models (CHMs) and arise from the non-linearities, can be rewritten as
\begin{equation}
\frac{v}{3 f^2} (h_2 \partial_\mu h_1 - h_1 \partial_\mu h_2) \partial^\mu h_2  =  \frac{v}{3 f^2} (s_\theta \, \partial_\mu h + c_\theta \, \partial_\mu H) (H \partial^\mu h - h \partial^\mu H)
\end{equation}
and give extra contributions to the interactions among the CP-even Higgs states. 
The  Feynman rules are 
\begin{align}
{[ h(p_1) h(p_2) h(p_3) ]} =& - i \lambda_{hhh},  \nn \\
{[ h(p_1) h(p_2) H(p_3) ]} =& - i \lambda^{(1)}_{hhH}   - i \lambda^{(2)}_{hhH} (p_1^2 + p_2^2 -2 p_3^2), 
\label{eq:trildef}
\end{align}
with 
\begin{align}
\lambda^{(2)}_{hhH} =  - v/(3 f^2) s_\theta.
\end{align}
The Yukawa Lagrangian, where we can directly read off the Feynman rules for the Higgs-fermion interactions, is given by
\begin{align}
    \mathcal L_\textrm{Yuk} = &- G_{h{\bar T}_{i}T_{j}}   \, \bar T_{Li} T_{Rj} h  - G_{H {\bar T}_{i}T_{j}}   \,  \bar T_{Li} T_{Rj} H   + \textrm{h.c.} \nn \\
&-  G_{hhT_iT_i}   \, \bar T_i T_i h^2  - G_{HHT_iT_i} \, \bar T_i T_i H^2 
 + \cdots, 
 \label{yuk}
\end{align}
with $i,j = 1, \ldots 9$, where, hereafter, $T_9$ will be used to denote the SM top quark after mixing with its 8 companions from compositeness. Furthermore, the 9 fermions are ordered with decreasing mass values and, as stated before, according to the experimental constraints, we enforced the bound 
\begin{align}
M_{T_8}>1.3 \mbox{ TeV.}
\end{align}
Note that there are no couplings of the $h$ and $H$ states with the exotic fermions with electric charge $Q=5/3$.  We furthermore assume negligible partial compositeness of the bottom quark (as well as of the light quarks and leptons). In this approximation, $h$ and $H$  do not couple to the bottom partners $B_{-1/3}$, introduced in Eq.~(\ref{eq:multi_dec}), and couple only to the top and to its heavy partners. In summary, 
since the latter interactions give the main contribution to the di-Higgs production cross section, in the following analysis, we will compare the C2HDM  with the SM by considering the fermion spectrum composed by the $Q=2/3$ states only, namely the $T_i$ states (with $i=1,..,9$).

\subsection{The Leading-Order Cross Section}
\label{sec:3}
We have derived the LO cross section for SM-like Higgs pair production following the conventions of \cite{Plehn:1996wb} and earlier works by us on Higgs pair production in the effective Lagrangian approach, respectively, CHMs (with only one Higgs boson, but partial compositeness in the top sector) \cite{Grober:2010yv,Gillioz:2012se,Grober:2015cwa,Grober:2016wmf}. 

The differential partonic cross section for $hh$ production can be
written in the form (for more details see App. \ref{sec:appendix1})
\begin{align}\label{eq:diffcxn}
\frac{d\hat{\sigma}(gg \to hh)}{d\hat{t}}=& \frac{\alpha_s^2}{512 (2\pi)^3} \nonumber \\
\times& \left[\left|\sum_{i=1}^{n_q} C^{hh}_{i,\triangle}
  F_\triangle (m_i) + \sum_{i=1}^{n_q} \sum_{j=1}^{n_q} \left(
    C^{hh}_{ij,\Box} F^{hh}_\Box 
    (m_i,m_j) + C^{hh}_{ij,\Box,5} F^{hh}_{\Box,5}(m_i,m_j \right) \right|^2 \right.
\nonumber \\
+& \left. 
    \left|\sum_{i=1}^{n_q} \sum_{j=1}^{n_q} \left( C^{hh}_{ij,\Box} G^{hh}_\Box
      (m_i,m_j) + C^{hh}_{ij,\Box,5} G^{hh}_{\Box,5} (m_i,m_j) \right)\right|^2 
\right] \;,
\end{align}
where $\alpha_s$ denotes the strong coupling constant and $m_{i,j}$ the mass of the quark $T_{i,j}$. The triangle and box form factors $F$ and $G$ are defined in Sect. \ref{app:formfactors} and $\hat{t}$ is defined in Eq. \eqref{eq:mandelstam}. We sum over the number
of quarks $T_{i}$ given in the model by $n_q=9$. The generalised couplings
$C^{hh}_{i,\triangle}$, $C^{hh}_{ij,\Box}$, and $C^{hh}_{ij,\Box,5}$ are defined as
\begin{eqnarray}\label{eq:CFactors1}
C^{hh}_{i,\triangle} &=& \frac{G_{h \bar{T}_iT_i} \lambda_{hhh}}{\hat{s}-m_{h}^2} 
+ \frac{G_{H \bar{T}_iT_i} \lambda_{Hhh}^{(1)}}{\hat{s}-m_{H}^2} 
+ \frac{G_{H \bar{T}_iT_i} \lambda_{hhH}^{(2)} (2m_h^2-2\hat{s})}{\hat{s}-m_{H}^2}
+ 2
  G_{hh\bar{T}_i T_i} \\
\label{eq:CFactors2}
C^{hh}_{ij,\Box} &=& g_{h\bar{T}_i T_j} \, g_{h\bar{T}_i T_j} \\
C^{hh}_{ij,\Box,5} &=& g_{h\bar{T}_i T_j,5} \, g_{h\bar{T}_jT_i,5} = - g_{h\bar{T}_i T_j,5} \, g_{h\bar{T}_iT_j,5} \;,
\label{eq:CFactors3}
\end{eqnarray}
where the $g_{h\bar{T}_i T_j}$ and $g_{h\bar{T}_i T_j,5}$ couplings are given by
\begin{eqnarray}
g_{h\bar{T}_i T_j} &=& \frac{1}{2} \left( G_{h \bar{T}_i T_j} + G_{h\bar{T}_j T_i} \right) \\
g_{h\bar{T}_i T_j,5} &=& \frac{1}{2} \left( G_{h \bar{T}_i T_j} - G_{h\bar{T}_j T_i} \right) \;,
\end{eqnarray}
and the partonic c.m.~energy squared $\hat{s}$ is defined in Eq. \eqref{eq:mandelstam}. 

The first three terms in Eq.~(\ref{eq:CFactors1}) stem from the second diagram and the last term from the third diagram in Fig.~\ref{fig:LODiagrams}, which involves the new quartic (2-fermion-2-scalar) coupling. The terms proportional to Eq.~(\ref{eq:CFactors2}) and (\ref{eq:CFactors3}) stem from the box, i.e.,~the first diagram in Fig.~\ref{fig:LODiagrams}, and give contributions corresponding to total gluon spin 0 (i.e.,~the form factors $F_\square^{hh}$, $F_{\square,5}^{hh}$) and 2 (i.e.,~the form factors $G_\square^{hh}$, $G_{\square,5}^{hh}$), respectively, along the collision axis. Note, in particular, that in the box
diagram the quark index does not have to be identical, i.e., different top partners can mix in the loop.

The above expressions were implemented into the code \texttt{HPAIR} \cite{HPair}, which was originally written for the computation of Higgs pair production in the SM and the MSSM. We modified \texttt{HPAIR} such that it allows us to calculate the inclusive cross section and the distributions in the Higgs pair invariant mass at LO QCD. Furthermore, the code was extended to also allow for the calculation of $p_T$ distributions (for the derivation, see App. \ref{app:pT}). For the calculation of the 1-loop integrals the program package \texttt{LoopTools} \cite{vanOldenborgh:1989wn,Hahn:1998yk} was used. The results for the inclusive cross section were checked by performing two independent calculations within our group using \texttt{HPAIR} and by cross-checking against \texttt{QCDLoop}~\cite{Carrazza:2016gav}. 

In order to smoothly cross the fermion thresholds in the loop diagrams, we have included in \texttt{HPAIR} the total widths of the $h$ and $H$ Higgs bosons in the $s$-channel propagators. In order to obtain the total widths of the Higgs bosons, we modified the Fortran code \texttt{HDECAY} \cite{Djouadi:1997yw,Djouadi:2018xqq} to calculate these and the BRs of the scalars $h$ and $H$ within the C2HDM. The code includes the state-of-the-art QCD corrections\footnote{Apart from the case mentioned here below, the QCD corrections can be translated from the SM and/or MSSM, with a minimum of effort, to the case of a 2HDM, whereas this is not possible for EW corrections.}  and the relevant off-shell decays. We further remark that the heavy Higgs boson can decay into a top quark and a top quark partner, however, since here no QCD corrections are available, these transitions are computed at LO. 
For the numerical analysis also the single Higgs production cross sections are needed. For this, the \texttt{FORTRAN} code \texttt{HIGLU} \cite{Spira_1995,spira1995higlu} was modified to calculate the C2HDM production of $h$ and $H$ through gluon fusion at LO in QCD. 

At this stage, a remark on Higher-Order (HO) corrections is in order. The QCD corrections to single and double Higgs production in the SM have been calculated and are known to be important. (For reviews on HO corrections to single and double Higgs production, see, e.g.,~\cite{LHCHiggsCrossSectionWorkingGroup:2011wcg,Dittmaier:2012vm,LHCHiggsCrossSectionWorkingGroup:2016ypw,Spira:2016ztx,DiMicco:2019ngk,Heinrich:2020ybq}.) Since we have additional quarks running in the loops, the QCD corrections cannot simply be taken over from the SM, however, not even in the heavy mass limit of the loop particle. We therefore have to restrict  ourselves in our analysis to the LO results. From previous calculations of NLO QCD corrections in UV complete models or EFT approaches \cite{Grober:2010yv,Gillioz:2012se,Nhung:2013lpa,Grober:2015cwa,Grober:2016wmf,Grober:2017gut,Abouabid:2021yvw,Borschensky:2022pfc,Arco:2022lai} in the heavy top mass limit as well as in the 2HDM including the full top mass dependence \cite{Baglio:2023euv}, we know that the SM-like Higgs pair production cross section at NLO QCD can be roughly approximated by adding a factor 2. For distributions, this is not possible, as here not only the overall normalisation changes, but also the shape. However, in this paper, we want to demonstrate what significant effects may arise in our BSM scenario: while higher-order QCD corrections may change the fine details of the findings, the overall coarse picture will not change.


\section{Numerical  Results \label{sec:numerical}}

\subsection{Input Parameters and Parameter Scan}
 For our numerical analysis, we use parameter sets that are compatible with the relevant theoretical and experimental constraints, as detailed in Sec.~\ref{sec:constraints}. 
In order to obtain these configurations, we performed a scan over  the parameter space of the various models we studied and kept only the compliant  points.

The fundamental parameters of the C2HDM are given by the compositeness scale, $f$, the gauge coupling of the underlying strong interaction, $g_\rho$, and the parameters of the fermionic sector as given in Eq.~(\ref{UV2}), namely, the Yukawa couplings of the heavy top quark partners $Y_{1,2}^{IJ}$ as well as the mixing between the latter and the 
elementary top quark $\Delta_{L,R}^I$ (which represents the leading contribution to the Coleman-Weinberg effective potential).

\begin{figure}
    \centering
    \includegraphics[scale=0.22]{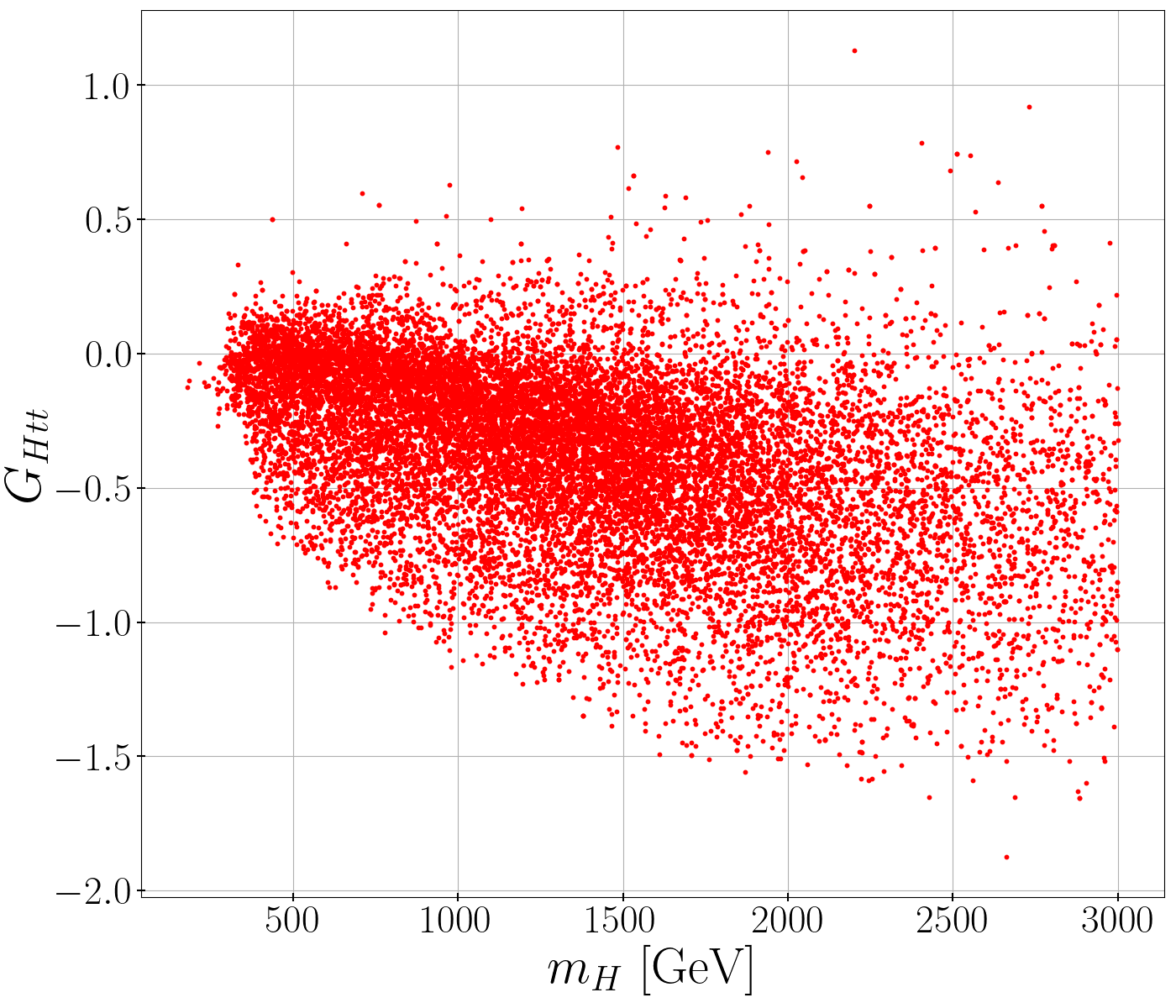}
    \includegraphics[scale=0.22]{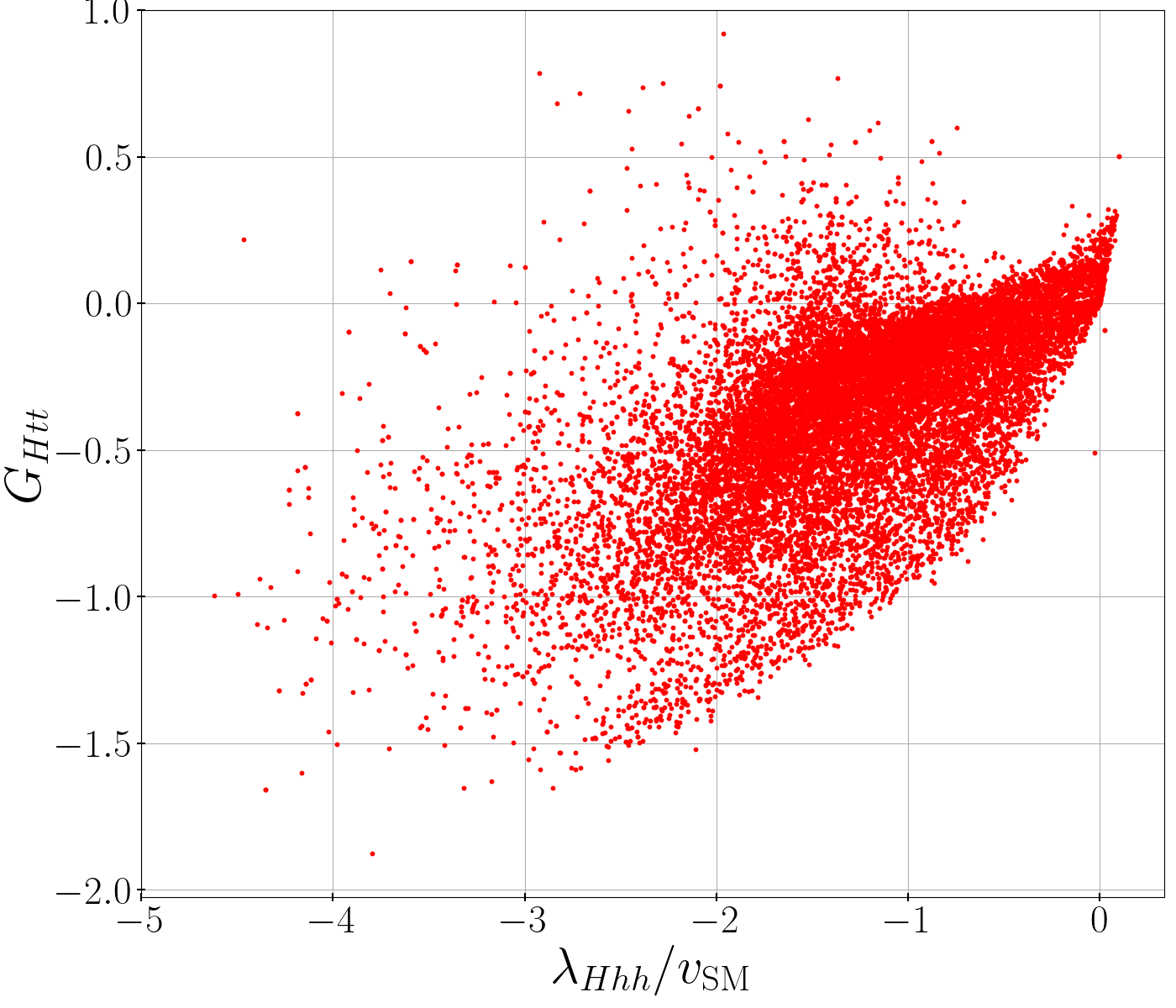}
    \caption{Correlations between  the Yukawa coupling $G_{Htt}$ and $m_H$ (left panel) as well as the trilinear coupling $\lambda_{Hhh}$ (right panel) in the C2HDM, where $\lambda_{Hhh} = \lambda_{hhH}^{(1)}$ as defined in Eq.~(\ref{eq:lscalarint}).}
    \label{correlation}
\end{figure}

The C2HDM parameter space has been explored by scanning the compositeness scale $f$ in the range (700, 3000) GeV, the strong coupling $g_\rho$ in the range (2,10) and the mixing $\Delta^I_{L,R}$ and Yukawa couplings $Y^{IJ}_{1,2}$  in the range $(-10,10)f$. 
The masses of  the heavier CP-even Higgs boson ($m_H$), the charged Higgs
boson ($m_{H^\pm}$), the CP-odd Higgs boson ($m_A$), the mixing angle $\theta$  between the two CP-even states ($h,H$, where by convention $h$ denotes the lighter one) and their couplings to fermions and bosons are all obtained from the aforementioned fundamental parameters. As  examples of this, correlations  between the Yukawa coupling $G_{Htt}$ and $m_H$ as well as between  $G_{Htt}$ and the trilinear coupling $\lambda_{Hhh}$ (normalised to $v_{\rm SM}$) are shown in Fig.~\ref{correlation}.

\begin{table}[]
    \centering
    \begin{tabular}{c|c}
       variable & value \\
       \hline 
       $\alpha_s(m_Z)$   & \SI{0.135}{} \\
      $\sqrt{s}$ & 13/14~TeV \\
      PDF  &  MMHT2014lo68cl  \cite{Harland_Lang_2015}\\
      ren. and fac. scale & $m_{HH} \slash 2 $ \\
    \end{tabular}
    \caption{List of input parameters used for  \texttt{HIGLU} and \texttt{HPAIR}. For details, see text.}
\label{tab:inputvalues}
\end{table}
The input values for \texttt{HIGLU} and \texttt{HPAIR}  for the generation of the leading-order single and di-Higgs cross sections, respectively, are listed in Tab.~\ref{tab:inputvalues}. 
For all of these, the mass values are computed from the C2HDM input values and vary from one parameter point to another\footnote{In the sample used for the numerical analysis, the top mass parameter lies between 165 and 175~GeV and the EW VEV $v_{\text{SM}}$ between 240 and 250~GeV.}. We take as input for the strong coupling constant, $\alpha_s$, its values measured at the $Z$ boson mass scale and evolve it to the relevant $\hat s$ value according to QCD, as threshold effects of the new heavy tops are negligible over their allowed mass ranges. We will use current single Higgs results to apply constraints from resonant di-Higgs searches on our di-Higgs results. This is why, in this case, the single Higgs production cross sections are computed at a c.m.~energy of 13~TeV. In contrast, the di-Higgs results are computed at 14~TeV, as we are here interested in the potential of the HL-LHC at the design energy. 
As we calculate only LO cross sections, we use LO Parton Distribution Functions (PDFs), with the renormalisation and factorisation scales both set to half the invariant Higgs pair mass $m_{HH}$. The value for $\alpha_s$ given in Tab.~\ref{tab:inputvalues} is taken from the chosen PDF.

\subsection{Application of di-Higgs Constraints \label{sec:dihiggscontraints}}
In addition to the constraints on the C2HDM, described in Sec.~\ref{sec:constraints}, we apply the constraints from di-Higgs searches on our parameter sample. The experiments give limits both on non-resonant and resonant searches. Since there is a smooth transition between these two cases in BSM Higgs pair production, which is not considered in the experimental analyses, we have to define ourselves criteria how to separate the two regimes, so that the experimental limits can be applied. We stress here, that we do not aim at a sophisticated application of experimental limits, which is best done by the experimental collaborations themselves, but rather take a pragmatic theory approach to make sure we do not consider scenarios that are clearly excluded by experiment. In our approach we follow the procedure outlined in \cite{Abouabid:2021yvw}, which we will briefly summarise in the following.\footnote{Applying slightly different criteria to separate the resonant and non-resonant regions may lead to the inclusion or rejection of different individual parameter points, which in an overall scan of the parameter space as done here should have no visible effects, however.}
\begin{figure}[tb]
    \centering
    \includegraphics[width=0.7\textwidth]{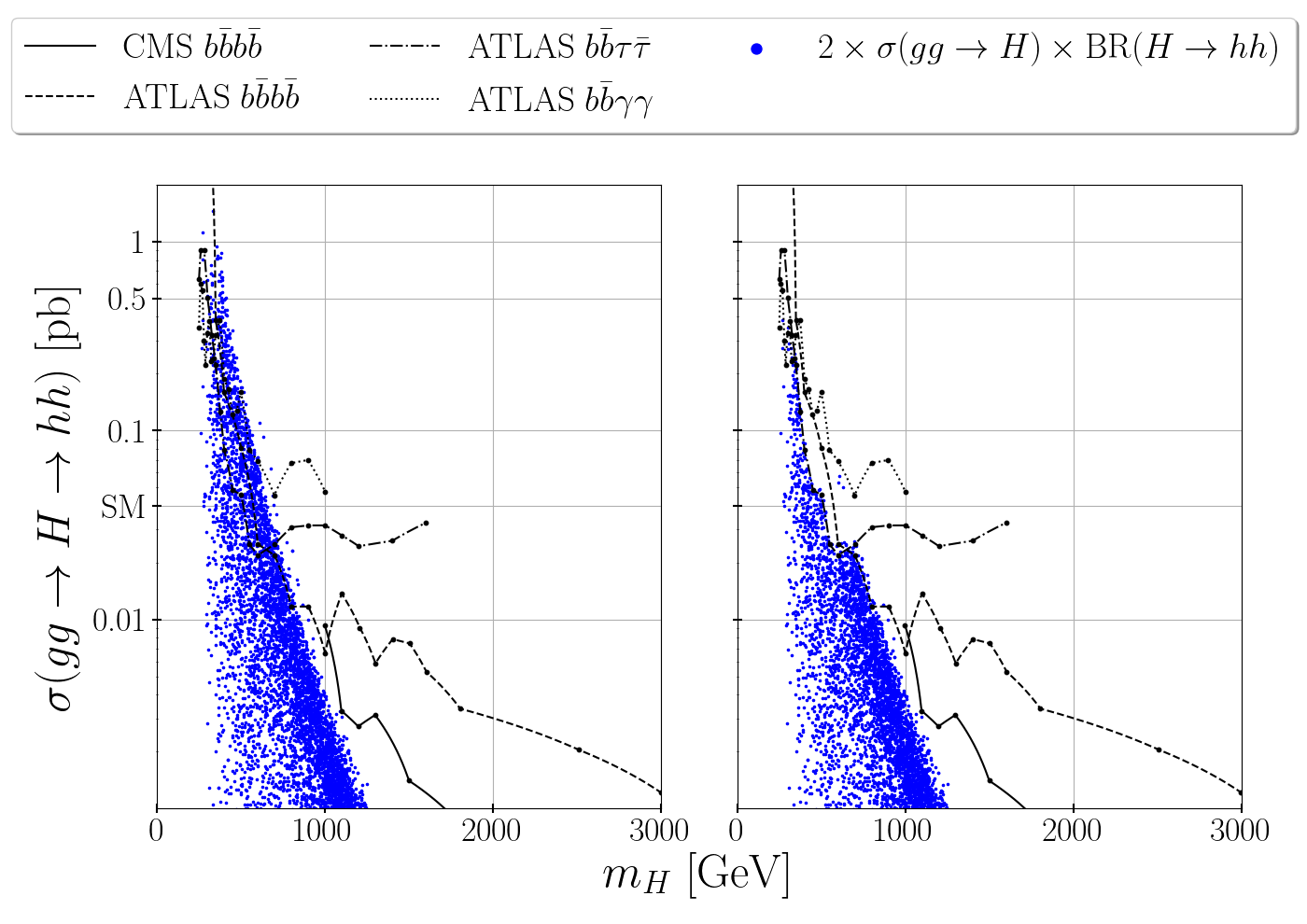}
    \caption{The resonant cross section (blue points calculated as  $2 \times \sigma^{\text{HIGLU}}(H)\times \mbox{BR}(H\to hh)$, where the factor 2 roughly accounts for the QCD corrections)  plotted against the heavy Higgs mass $m_H$. Left (right) without (with) the constraint from the experimental resonant di-Higgs searches in the various final states \cite{CMS:2021qvd,ATLAS:2021ifb,ATLAS:2022hwc,ATLAS:2022xzm} as indicated in the legend. The value of the SM cross section, $\sigma_{hh}= 2 \times 19.96$~fb, is indicated as well.}
    \label{fig:Resonantconstraints}
\end{figure}

A pair of SM-like Higgs bosons $hh$ is clearly produced \textit{non-resonantly} if the mass $m_H$ of the heavier Higgs boson is below the $hh$ threshold, i.e.,~$m_H < 2 m_h$. However, also in cases where resonant production is possible, it may be very suppressed. The reason can be $s$-channel suppression due to a large mass $m_H$ and/or a large total width $\Gamma_H$, small $H$ couplings (top-Yukawa coupling and/or $\lambda_{Hhh}$), or destructive interferences between different diagrams. From an experimental point of view, the total cross section would not be distinguishable from truly non-resonant production then. We therefore define a cross section to be non-resonant when the single heavy $H$ production with subsequent decay into SM-like Higgs bosons $hh$ makes up for less than 10\% of the total di-Higgs cross section, and accordingly apply the non-resonant limits. The limits from resonant searches, however, we always apply whenever $m_H > 2 m_h$ (apart from the exception detailed in the next paragraph).

For the application of the \textit{resonant} constraints, we compute the heavy $H$ production  cross section with \texttt{HIGLU} and multiply it with the BR of the $H \to hh$ decay obtained from \texttt{HDECAY}. To roughly account for the QCD corrections, we additionally include a factor two. The resulting rate is multiplied with the BRs of the SM-like Higgs $h$ into the various final states (again obtained from \texttt{HDECAY}), where the resonant searches have been performed \cite{ATLAS:2018rnh,ATLAS:2018uni,PhysRevLett.122.089901,ATLAS:2020azv,ATLAS:2018dpp,ATLAS:2018fpd,CMS:2020jeo,ATLAS:2018hqk,ATLAS:2018ili,CMS:2021qvd,ATLAS:2021ifb,ATLAS:2022hwc,ATLAS:2022xzm}. Points exceeding at least one of these limits are rejected. Note, that we apply this procedure only to points where the total width $\Gamma_H$ of the heavy Higgs does not exceed 5\% of its mass values, i.e.,~$\Gamma_H/m_H \le 5$\%. This allows us to roughly account for the fact that the experiments apply the narrow-width approximation. Otherwise, no resonant limits are applied, and the point is kept in the sample.

We show in Fig. \ref{fig:Resonantconstraints} (left) the calculated resonant production cross sections of all BPs before applying the limits, whereas the right plot shows the situation after the application of all above mentioned limits from the experimental searches. For illustration, we have included the limits of some of the searches, namely those from \cite{CMS:2021qvd,ATLAS:2021ifb,ATLAS:2022hwc,ATLAS:2022xzm}, which are the most stringent ones. The comparison with the right plot after application of the limits shows that the resonant searches are already sensitive to our model and  constrain parts of its parameter space. Note that the remaining points above the experimental limits are there because they do not fulfil the narrow width approximation. 

\subsection{Impact of New C2HDM Effects on Higgs Pair Production}
Before we turn to the detailed discussion of resonant and non-resonant Higgs pair production in our model, we analyse the impact that the new effects emerging in the C2HDM have on the inclusive Higgs pair production cross section.

\begin{figure}[tb]
    \centering
    \includegraphics[width=0.6\textwidth]{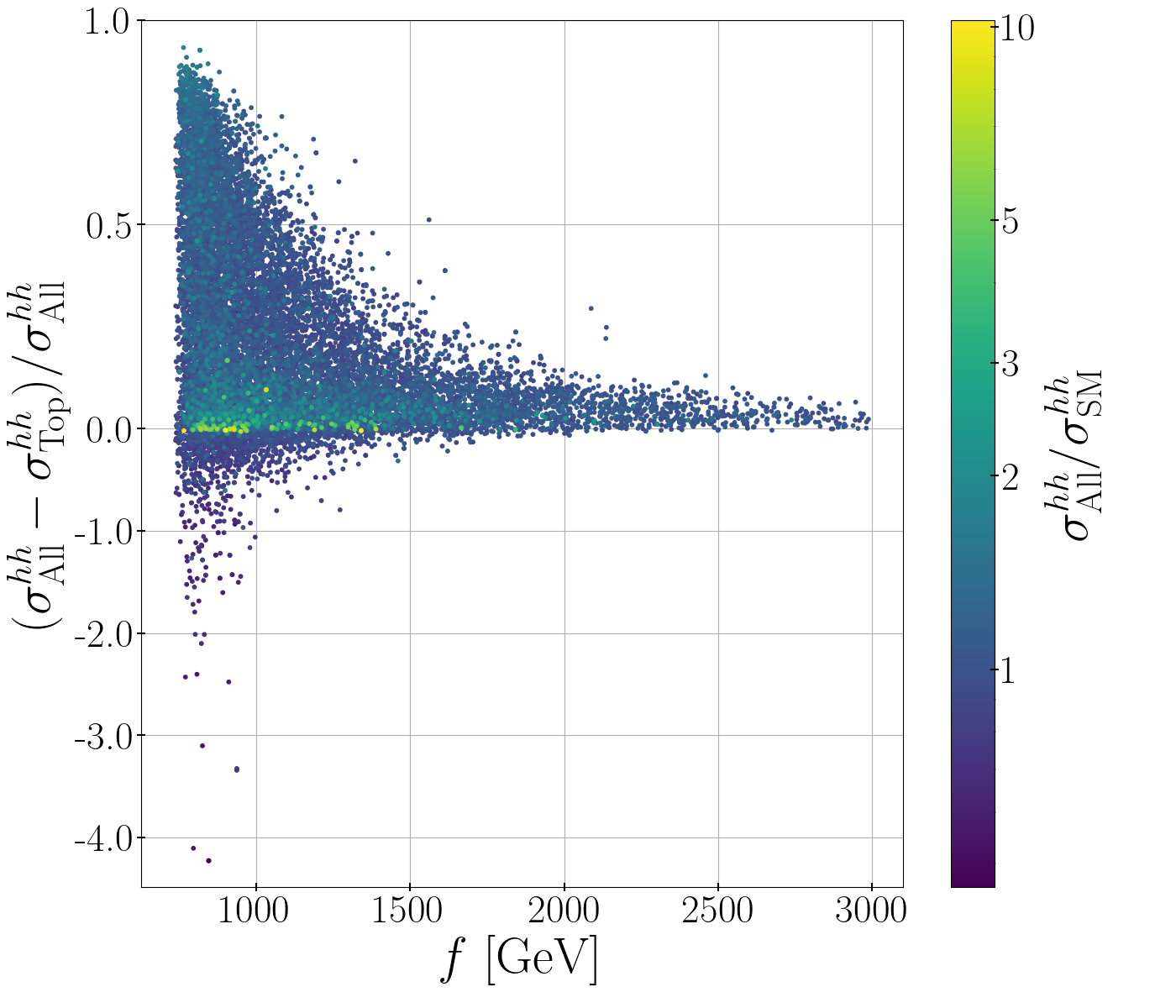}
    \caption{The full cross section minus the cross section obtained using only the top quark in the loops, normalised to the full cross section, plotted against the compositeness scale $f$. The colour code displays the ratio of the full cross section and the SM cross section. Note that for illustrative reasons, we rescaled the negative $y$-axis.}
    \label{fig:RelativeDifferenceheavyQuarks}
\end{figure}

\begin{figure}[tb]
    \centering
    \includegraphics[width=0.6\textwidth]{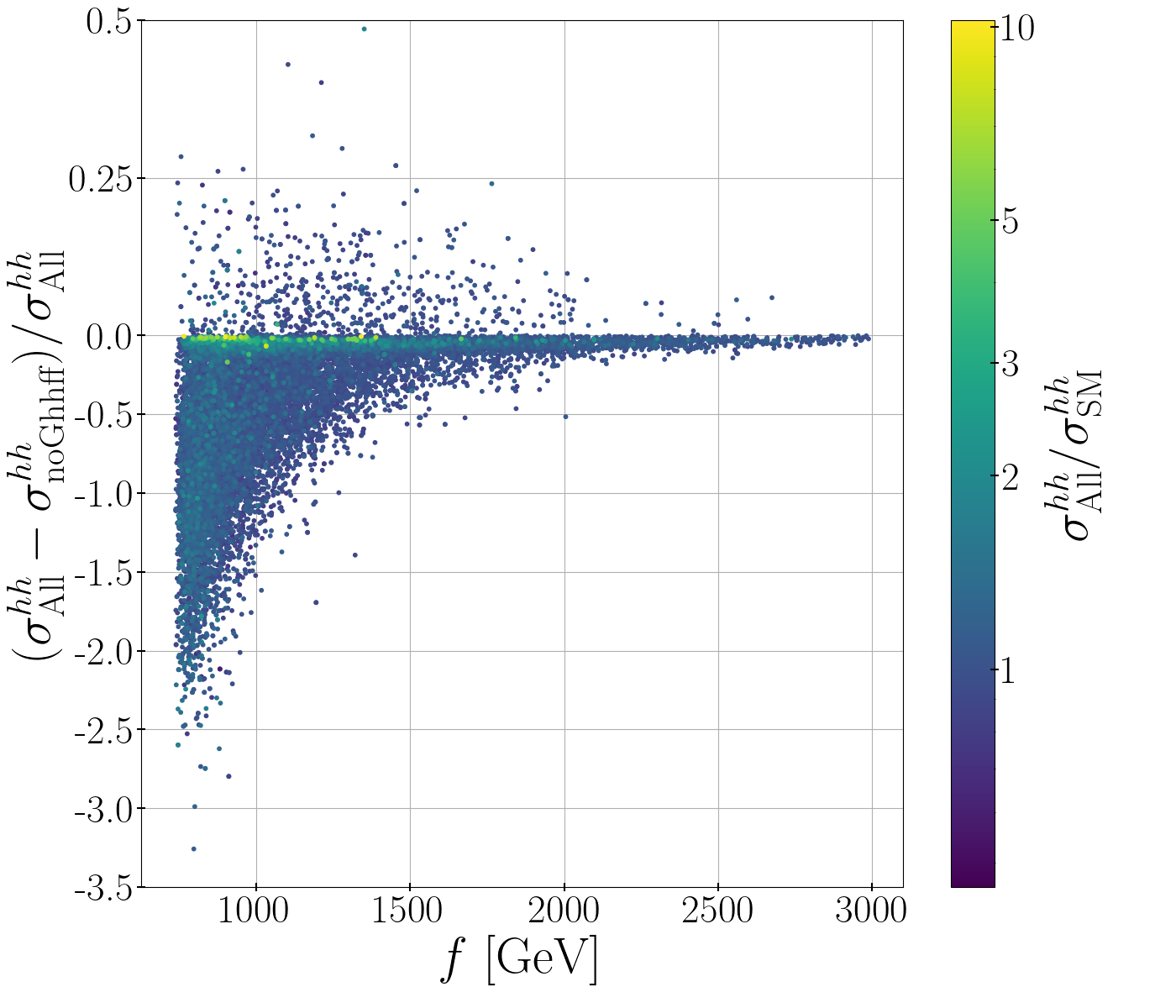}
    \caption{The full cross section minus the cross section obtained neglecting the quartic scalar-scalar-fermion-fermion couplings, normalised to the full cross section. The colour code is the same as in Fig.~\ref{fig:RelativeDifferenceheavyQuarks}, and again for illustrative reasons, we rescaled the negative $y$-axis.}
    \label{fig:RelativeDifferencenoGhhff}
\end{figure}

In Fig. \ref{fig:RelativeDifferenceheavyQuarks} we show the impact of the heavy top partners on the production cross section of a pair of SM-like Higgs bosons, by plotting the difference between the full cross section and the cross section where we only consider the top quark running in the loop, normalised to the full cross sections, as a function of the compositeness scale $f$. The colour code indicates the ratio of the full cross section in our model and the SM cross section. All cross sections are obtained at LO. Since we investigate ratios, this practically does not change the results. First of all it is evident, that the heavy top partners can interfere constructively or destructively and thereby enhance or suppress the overall cross section. Furthermore, with increasing compositeness scale $f$ the relative contribution of the heavy top partners diminishes, i.e.,~the composite sector decouples. Moreover, it can be seen that for points with a large total cross section the relative impact of the heavy top partners is small. These large cross section values (yellow and orange points) are due to resonant heavy Higgs production. This behaviour is to be expected since in the resonant case, the triangle diagram (see Fig. \ref{fig:LODiagrams} (middle)) is the dominant one and here the heavy top partner contributions are suppressed by their masses. Therefore, the main contributions come from the diagrams with at least one (light) top quark in the loop. For small $f$ and di-Higgs cross sections between 1/2 and 2 times the SM value, the relative contributions of the heavy top partners to the total cross section can be substantial, however.

We also analysed the effect of the novel quartic scalar-scalar-fermion-fermion coupling by displaying in Fig. \ref{fig:RelativeDifferencenoGhhff} the full cross section minus the cross section without the contribution from the quartic coupling normalised to the full cross section, as a function of the compositeness scale. The colour code is the same as in Fig. \ref{fig:RelativeDifferenceheavyQuarks}.  Similar to the results regarding the impact of the heavy quarks, in the limit of a large compositeness  scale $f$ we see again the decoupling of the composite sector. And in the resonant case, i.e.,~where we have large di-Higgs cross sections, the quartic coupling contribution cannot compete with the resonant effect either. For parameter scenarios, however, where the resonant contribution is zero or subdominant\footnote{Even if kinematically resonant production is possible, it can be suppressed due to the above mentioned effects.}, 
the effective coupling can have a large impact on the overall cross section, both in a constructive or a destructive way. We see a tendency towards larger destructive than constructive interferences in the plot, which corresponds to predominantly positive values for the novel quartic coupling, as the diagram with the quartic coupling (Fig.~\ref{fig:LODiagrams} (right)) interferes constructively with the remaining triangle diagrams (Fig.~\ref{fig:LODiagrams} (middle)) if the quartic coupling is positive. This in turn then leads to a larger destructive interference between the total triangle and the box contribution, so that for positive quartic couplings we encounter destructive interferences.

\section{The Non-Resonant Case \label{sec:nonresonant}}
\begin{figure}[tb]
    \centering
    \includegraphics[width=0.6\textwidth]{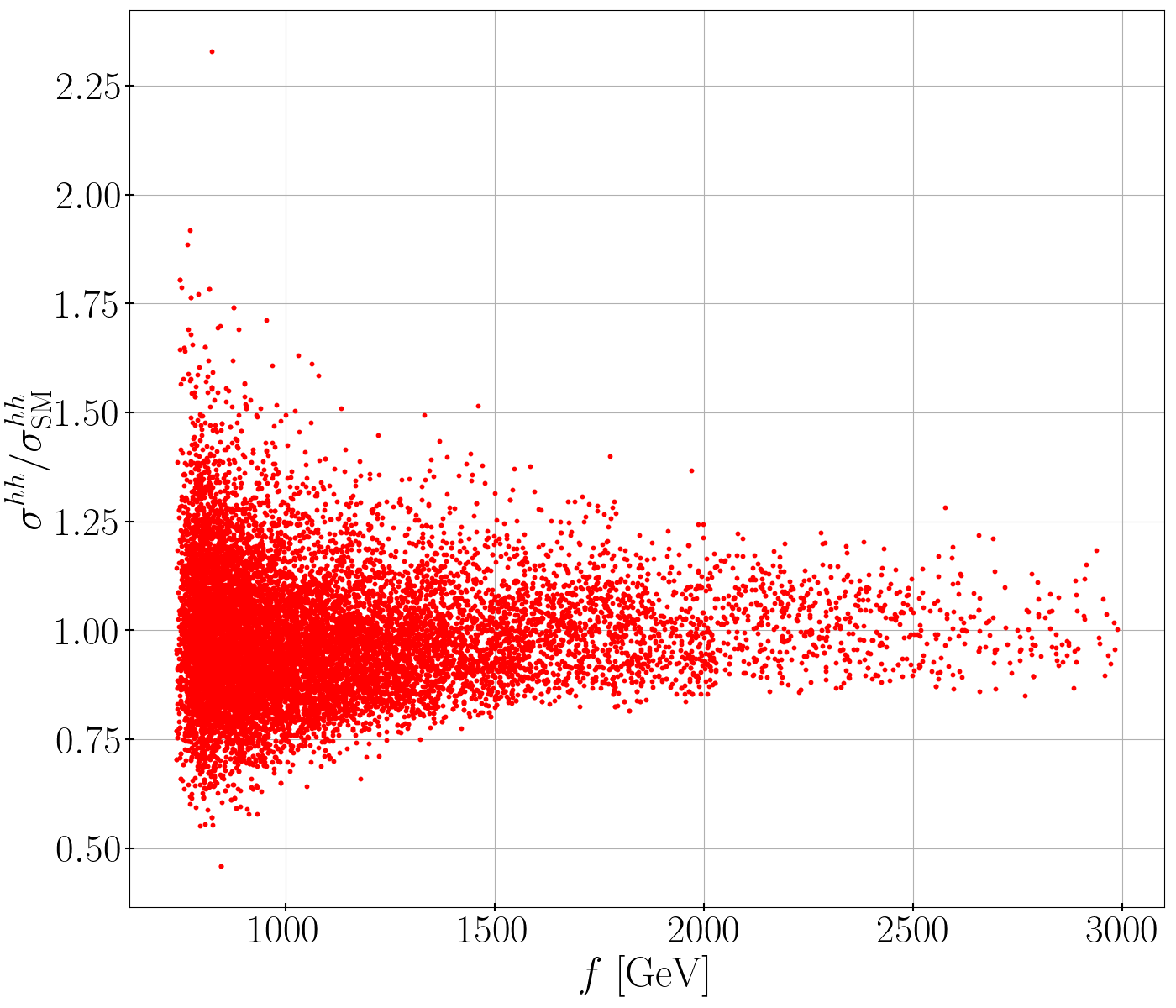}
    \caption{The full cross section of non-resonant points normalised to the SM cross section plotted against the compositeness scale $f$.}
    \label{fig:inclusivecxn_non-resonant}
\end{figure}

In this section, we discuss the case where the cross section is not dominated by resonant production. As detailed in Subsec.~\ref{sec:dihiggscontraints}, we include in the non-resonant case  not only scenarios where $m_H < 2 m_h$ but also such scenarios where the resonant production cross section (in the narrow width approximation) remains below 10\% of the full inclusive cross section, i.e.,
\begin{align}
\frac{\sigma(gg \to H) \times \mathrm{BR}(H \to hh)}{\sigma(gg \to hh)} <0.1 \;. \label{eq:nonresconstr}
\end{align}
Note, however, that on these points with (though small) resonant contributions we also applied the resonant limits to be sure they are not excluded already by the resonance searches.
We will first discuss the inclusive results and then move on to the differential distributions, investigate the effect of the different couplings and study the resulting interference patterns.

\subsection{Inclusive Results}
In Fig.~\ref{fig:inclusivecxn_non-resonant} we display, as a function of the compositeness scale $f$ the LO $hh$ cross section of our model normalised to the SM value for the non-resonant points, which we obtained from our sample of allowed parameter points after applying the condition Eq.~(\ref{eq:nonresconstr}). For small compositeness scales, non-resonant production in our model can be larger by up to a factor of about 2.3 times the SM value, it can, however, also be suppressed by up to a factor of about 0.5 the SM value. The current limit from a recent ATLAS study combining several final states lies at 2.4 the SM values at 95\% confidence level \cite{ATLAS:2022jtk}, so that the model starts to be tested also by the non-resonant searches.

\subsection{Exclusive Results}
\begin{figure}[tb]
    \centering
    \includegraphics[width=0.6\textwidth]{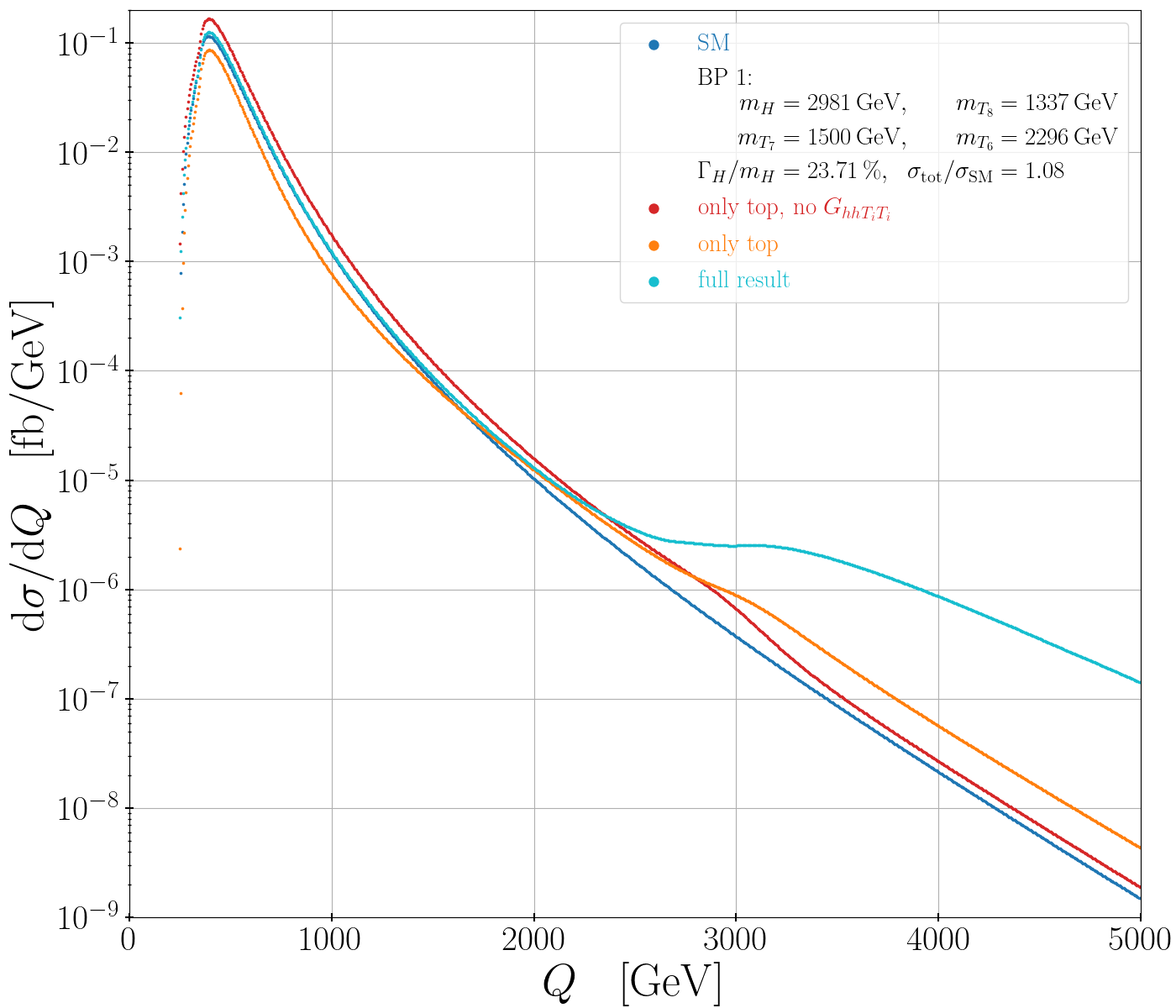}
    \caption{Invariant mass distribution for BP 1. Blue: SM result; light blue: full C2HDM result; orange: C2HDM including only top loops; red: C2HDM with only top loops and no quartic 2-Higgs-2-fermion coupling. We also indicate the heavy Higgs mass and total width to mass ratio and the masses of the three lightest heavy top quarks, as well as the ratio of the total inclusive C2HDM cross section to the SM value.}
    \label{fig:invMassBP4423signnew}
\end{figure}

\begin{figure}[tb]
    \centering
    \includegraphics[width=0.6\textwidth]{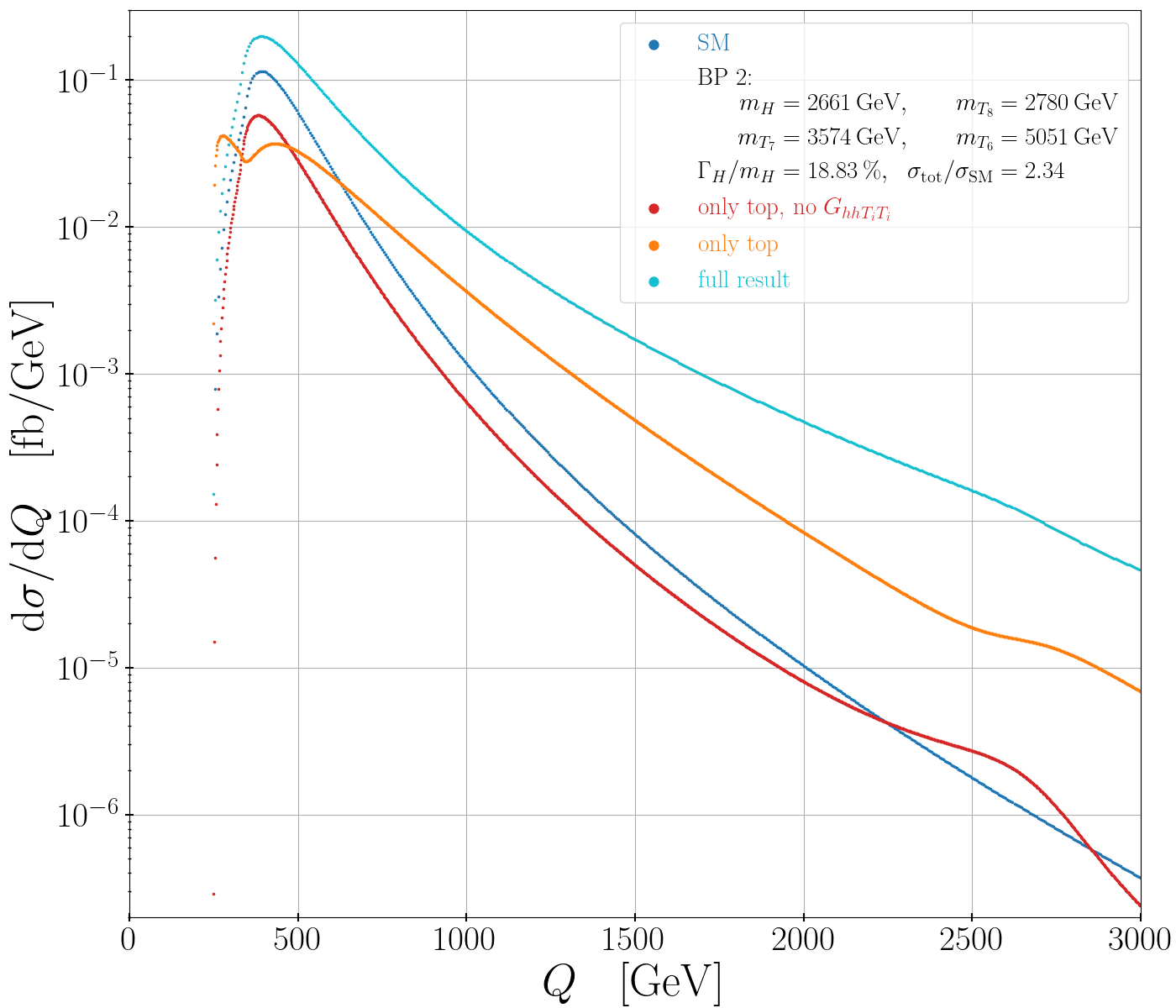}
    \caption{Invariant mass distribution for BP 2. Colour code as in Fig.~\ref{fig:invMassBP4423signnew}.}
    \label{fig:invMassBP1901sign}
\end{figure}

For the discussion of the exclusive results, we selected BPs displaying some interesting features.\footnote{The input values of the discussed BPs are summarised in Appendix~\ref{app:bps}.} 

In Fig. \ref{fig:invMassBP4423signnew} we show the invariant mass ($Q\equiv m_{hh}$) distribution for a BP with a heavy Higgs boson mass of 2.98~TeV. It is non-resonant due to its negligible resonant contribution to the total cross section. However, the distribution will show the resonance. Let us first note, that for this BP neglecting the heavy quarks, i.e., only taking into account  diagrams with a top quark in the loop (orange line), results in a distribution that is below the SM expectation (blue line), until about \SI{1.6}{\tera \electronvolt}. On the other hand, the distribution obtained by additionally neglecting the quartic scalar-scalar-fermion-fermion coupling (red) is always above the SM expectation. We can hence clearly see for this point the destructive interference due to the novel quartic coupling. 
If we look then at the distribution for the  full cross section including all diagrams (light blue), we are back at around the SM expectation, below the onset of the heavy quark effects at the threshold of $2m_{T_8} \approx 2.6$~TeV and the heavy Higgs resonance effect at $m_H \approx 3$~TeV. So, the constructive interference of the  heavy quarks is zeroed by the negative effect due to the quartic coupling. Note also, that the effect of the heavy quarks is more important than the resonance effect. This feature, if detected, clearly indicates the presence of new physics in di-Higgs production.

The distributions of another interesting BP are displayed in Fig. \ref{fig:invMassBP1901sign}. Here we have a large quartic coupling inducing a destructive interference with the other contributions, as can be seen from the kink in the orange line at around 300~GeV in comparison to the red line. The quartic contribution also dominates for high energies. This is to be expected for large enough quartic couplings, since in the quartic coupling diagram there is no additional scalar propagator (compared to the other triangle diagrams) so that it is not as suppressed at large invariant masses. 
This explains, why already for the differential distribution involving only the top quark (orange line) the shape differs significantly with respect to the SM (blue line) for this BP. Including now also the heavy quarks (light blue line), we see that overall we can have a significant enhancement of the cross section with respect to the SM result. Here then also the impact of the heavy quarks plays a role and the interference between all contributions. And finally, we also see a small kink at around \SI{2.7}{\tera \electronvolt} indicating the resonance at the heavy Higgs.

\section{The Resonant Case \label{sec:resonant}}
We now turn to the discussion of the resonant case, both inclusively and exclusively. Note that here we discuss BPs which would not be labelled 'resonant' according to the definition given in Sec.~\ref{sec:dihiggscontraints}, but they exhibit a clear resonance peak in the distributions.

\subsection{Inclusive Results}
\begin{figure}[ht!]
    \centering
    \includegraphics[width=0.6\textwidth]{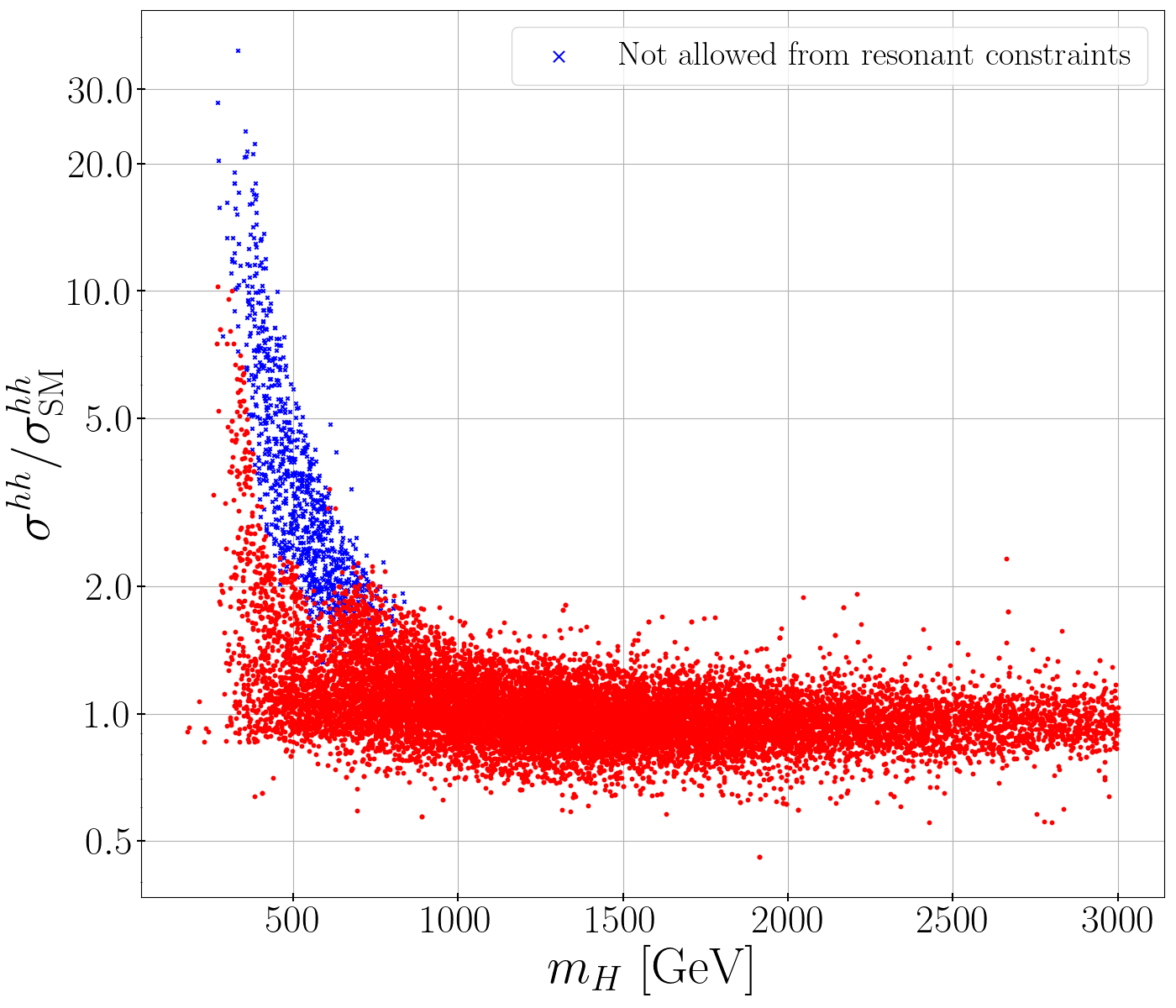}
    \caption{The total cross section normalised to the SM result, plotted against $m_H$, where the blue points are excluded by resonant searches.} 
    \label{fig:fsigtotupdated}
\end{figure}

The parameter points displayed in Fig.~\ref{fig:fsigtotupdated} exhibit resonant SM-like di-Higgs production. Red points fulfil all constraints, blue points are excluded by the resonant di-Higgs searches at the LHC. We show the inclusive cross section normalised to the SM value as a function of $m_H$. 
First of all, we see that we can have a significant enhancement of the cross section, by up to more than 30 times the SM cross section. Such large cross sections are, however, excluded (blue points) by the resonant di-Higgs searches at the LHC. The explored parameter region is hence already sensitive to resonant searches. The maximal cross section allowed by experiment that was found in our sample, is around 10 times the SM cross section (similar to other models with resonant production, see, e.g.,~\cite{Abouabid:2021yvw}). Furthermore, as can be inferred from the plot, with increasing $m_H$ (which corresponds to increasing $f$ as $m_H$ grows with $f$ for large $f$ \cite{DeCurtis:2018zvh}) we converge back to the SM result. 

\begin{figure}[h!]
    \centering
    \includegraphics[width=0.6\textwidth]{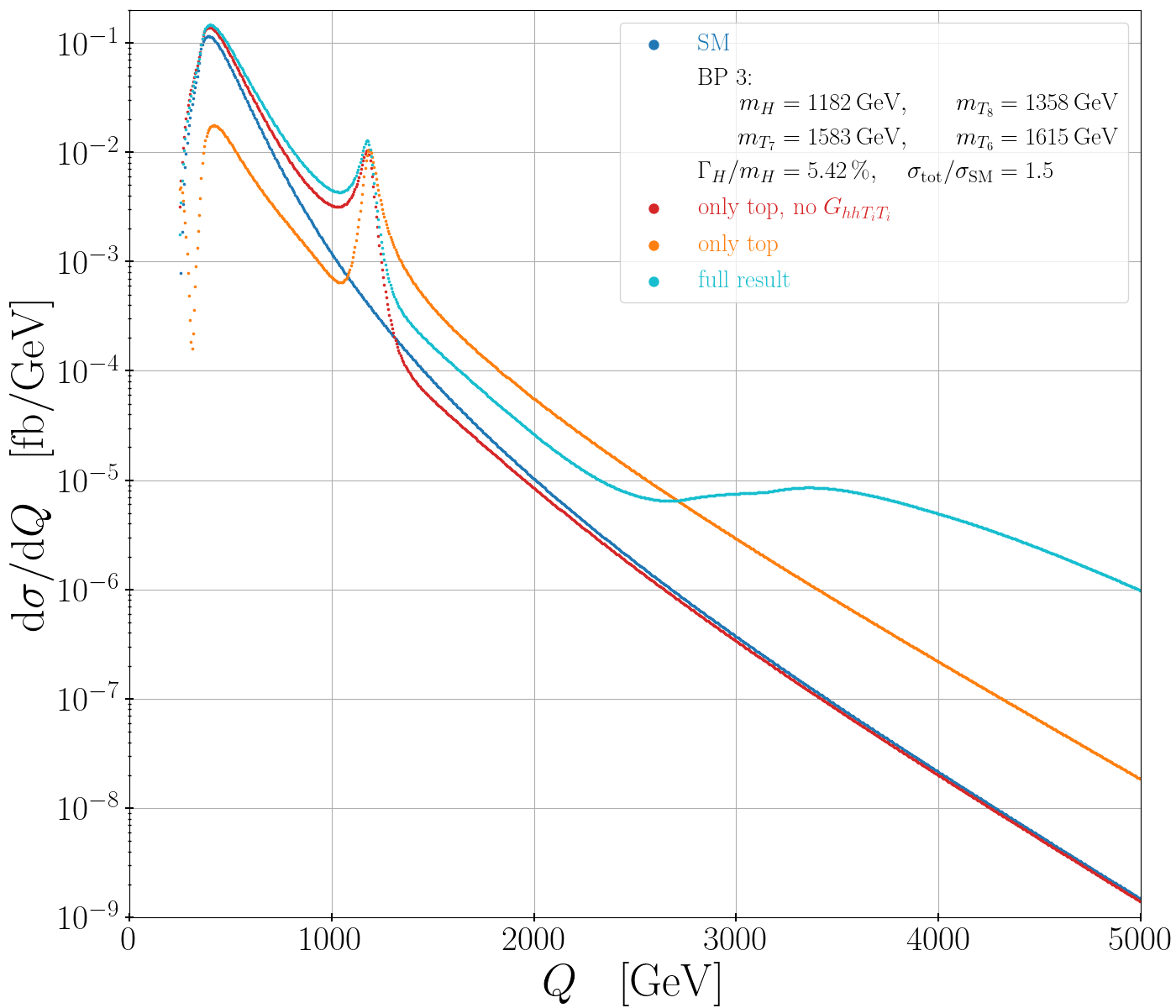}
    \caption{Invariant mass distribution for the BP 3. Colour code as in Fig.~\ref{fig:invMassBP4423signnew}.}
    \label{fig:invMassBP0687signnew}
\end{figure}

\subsection{Exclusive Results}

\begin{figure}[h]
    \centering
\includegraphics[width=0.6\textwidth]{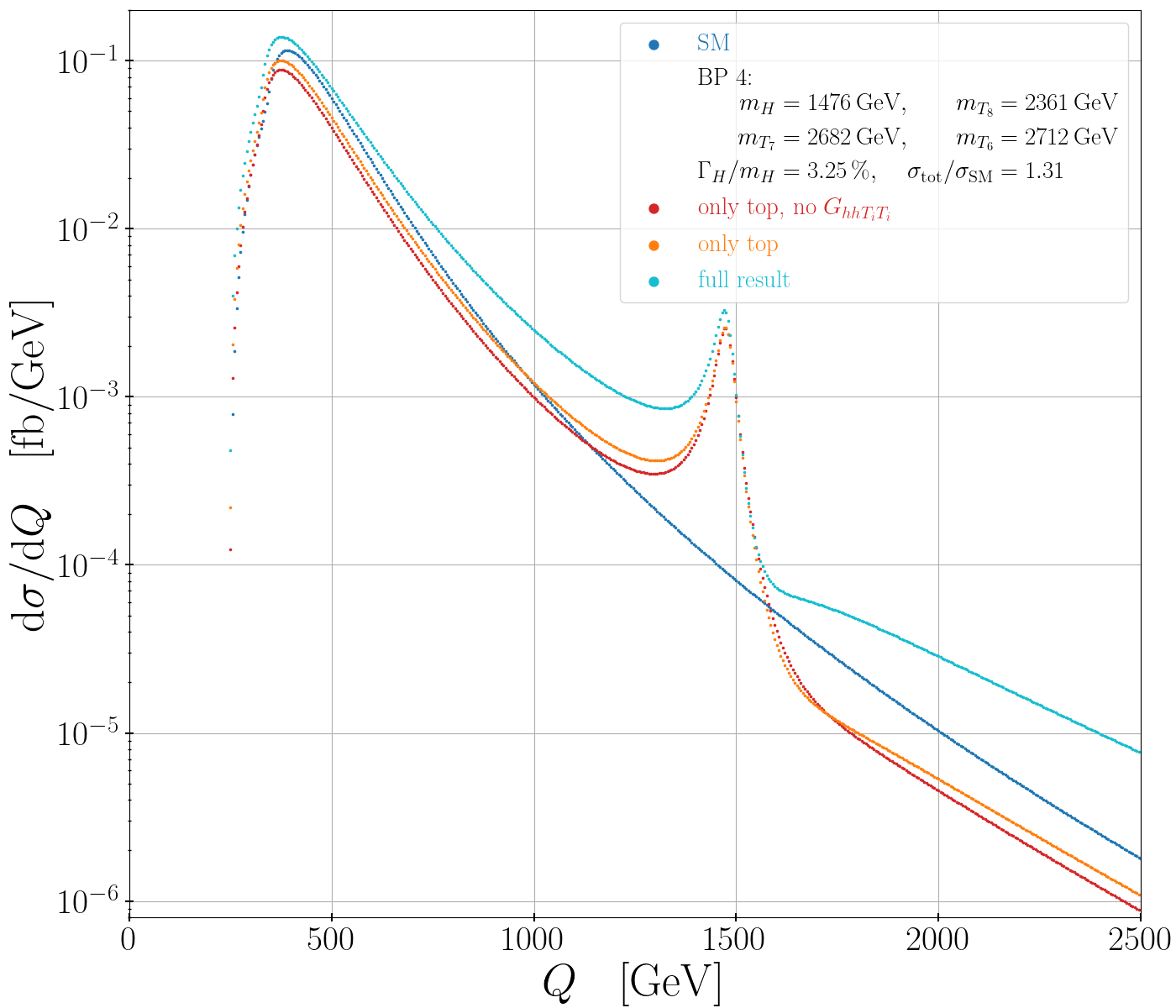}
    \caption{Invariant mass distribution for BP 4. Colour code as in Fig.~\ref{fig:invMassBP4423signnew}.}
    \label{fig:invMassBP2856signnew}
\end{figure}

\begin{figure}[tb]
    \centering
    \includegraphics[width=0.6\textwidth]{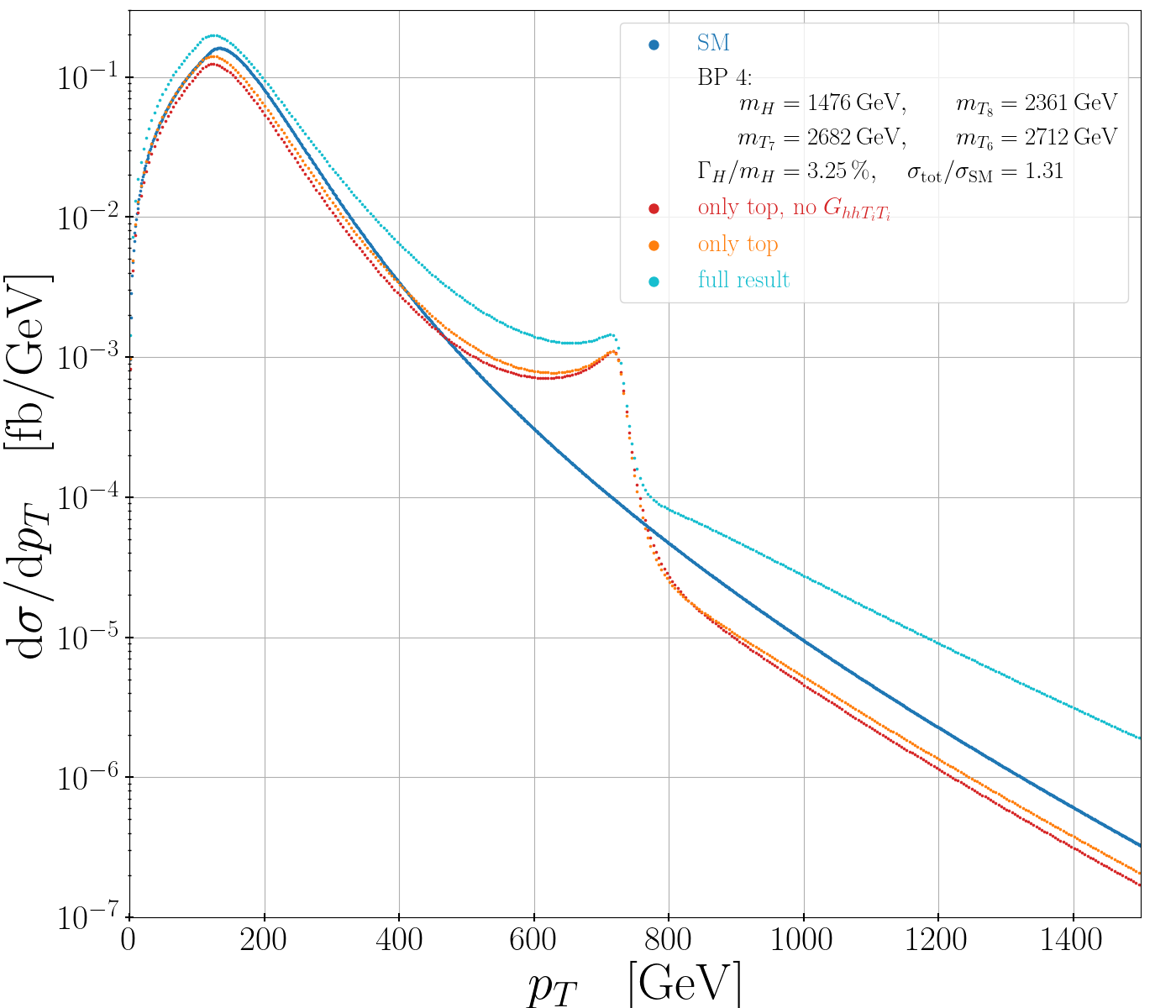}
    \caption{Differential $p_T$ distribution of the BP 4. Colour code as in Fig.~\ref{fig:invMassBP4423signnew}.}
    \label{fig:pTBP2856sign}
\end{figure}

In Fig \ref{fig:invMassBP0687signnew} we can see a BP with a clear resonance at the heavy Higgs mass $m_H=1.2$~TeV and also a nicely visible threshold effect for the heavy top partners setting in at $Q \approx 2.7$~TeV. Due to the relatively large Higgs mass, the effect of the resonance contribution on the total cross section is rather small so that we obtain a cross section value of the order of the SM expectation. The shape of the differential cross section, however, is quite interesting since we have several contributions, stemming from the heavy resonance, the heavy top partners and from the quartic scalar-scalar-fermion-fermion couplings, which all can interfere with each other. Let us have a closer look at the various effects and investigate the difference of shapes in our model compared to an elementary 2HDM. In the plot, the elementary 2HDM is mimicked by the red curve when the heavy quarks and the quartic 2-Higgs-2-fermion couplings are turned off. We then see that, in this example, the full cross section of our model (light blue line) is enhanced compared to the SM value (blue line) both before and after the peak in contrast to the elementary 2HDM-like result. In the latter case there is only the triangle diagram with the heavy Higgs which interferes with the SM-like triangle and box diagrams. The interference term is proportional to $(Q^2-m_H^2)$ and hence changes sign before and after the resonance, which explains the observed behaviour. When we additionally add in the contribution of the diagram with the quartic 2-Higgs-2-top quark coupling (orange line), which in this example interferes destructively as the coupling has a positive sign, the effect is inverted, we have a suppression before the resonance and an enhancement after the resonance compared to the SM result. Adding in finally also the heavy quarks we have our full model (light blue line) and their contribution finally leads to the described behaviour of the distribution. Thus, in principle, the interference patterns around a resonance can be used to distinguish between an elementary 2HDM and a composite 2HDM.

A similar behaviour can also be seen in Fig. \ref{fig:invMassBP2856signnew} where we again see an enhancement of the shape of the full cross section (light blue) compared to the SM (blue) before and after the resonance, and a clear difference between the full result and the SM result or the result obtained without heavy quarks (orange) and the elementary 2HDM-like result (red). For this BP we also calculated the differential $p_T$ distribution. The derivation of the distribution is given in App. \ref{app:pT}, which we implemented in \texttt{HPAIR}. The resulting distribution is depicted in Fig. \ref{fig:pTBP2856sign}. Similar features as in the invariant mass distributions can be seen, but of course the resonance here is not at the mass of the heavy Higgs. Thus, the same conclusion as in the invariant mass distribution can be reached, i.e., the shape of the distribution around a resonance is clearly changed by interference effects with additional contributions from both heavy quarks and quartic couplings.

\subsection{Binned Distributions}
\begin{figure}[tb]
    \centering
     \includegraphics[width=0.6\textwidth]{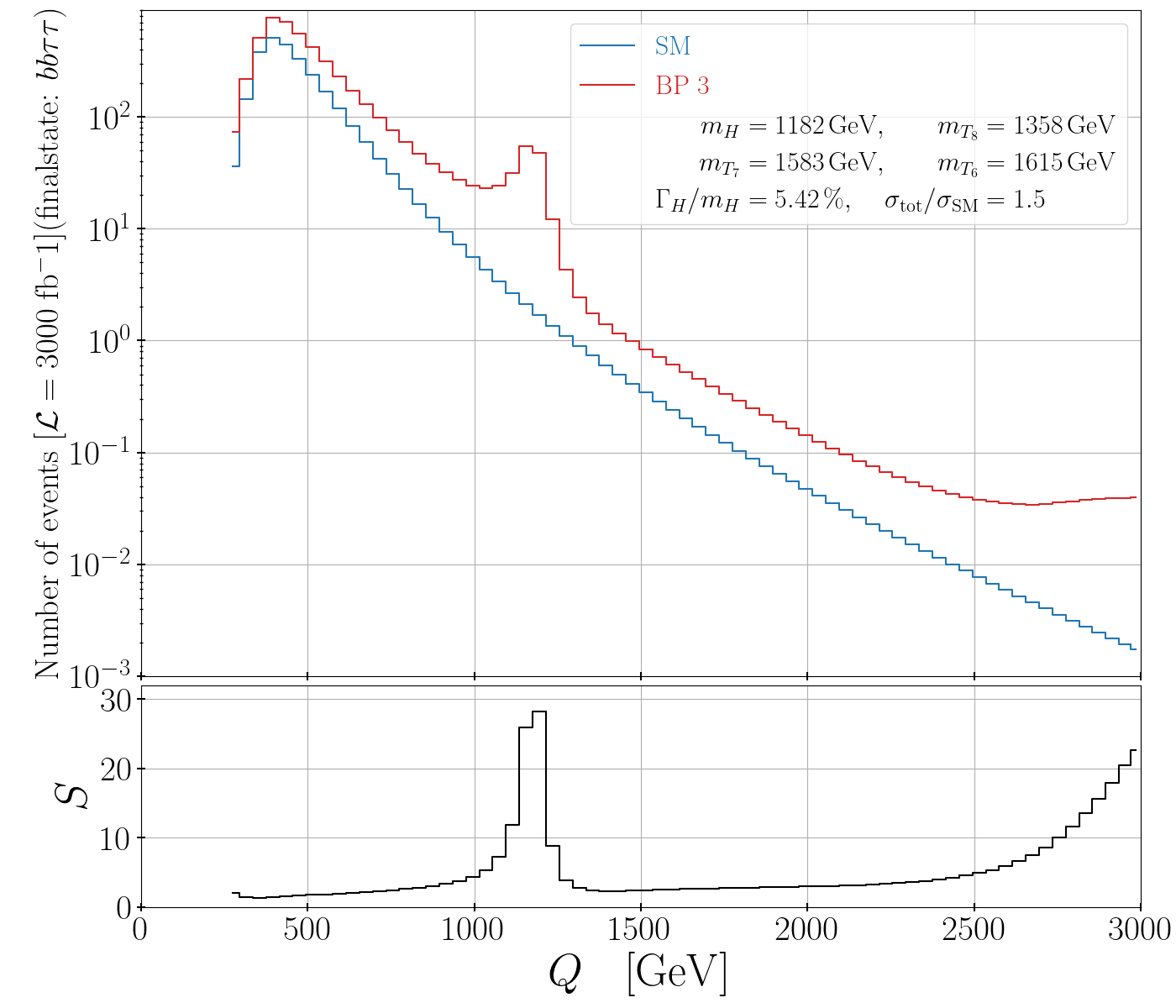}
    \caption{Upper: Binned distribution of the number of events $N_{\text{BP}}$, for a chosen bin size of 40~GeV and an integrated luminosity of $\int {\cal L}=3000$~fb$^{-1}$ in the $bb\tau\tau$ final state for the chosen BP (red) and the SM expectation (blue). Lower: Ratio $S$ of number of events per bin w.r.t.~to the corresponding number of events in the SM. }
    \label{fig:binneddistributionBP0687sign}
\end{figure}
We also analysed for selected  BPs in a simplified way the expected number of signal events by roughly simulating the experimental environment. We used the obtained differential distributions and calculated the expected number of events, given a specified bin range, luminosity and final state. We then compared the expected events $N_{\mathrm{BP}}$ for a given BP of our model with the SM expectation $N_{\mathrm{SM}}$ bin by bin by calculating the ratio
\begin{align}
S= \frac{N_{\mathrm{BP}}}{N_{\mathrm{SM}}}.
\end{align}
In our example we selected a bin width of \SI{40}{\giga \electronvolt}, an integrated luminosity of \SI{3000}{\per\femto \barn} and the $bb\tau \tau$ final state, i.e., we multiplied the results with the respective BRs (the SM with the SM BRs and our BP with the calculated Higgs BRs). It is clear that this is only a naive estimate where a lot of effects from both theory (e.g., NLO corrections to differential distributions\footnote{The ratios of the total C2HDM and SM cross sections remain roughly unchanged when going to NLO QCD, as results in several different BSM models showed, see e.g.~\cite{Abouabid:2021yvw}. This is not the case, however, for differential distributions where the shapes change as well when including the NLO QCD corrections. Since unfortunately they are not available for our model, we have to take the LO results.}, etc.) and experiment (e.g., acceptance and selection cuts, reducible and irreducible backgrounds, etc.) are not included. Thus,  this should only be taken as a first rough estimate under optimal conditions. The corresponding plot is displayed in Fig.~\ref{fig:binneddistributionBP0687sign}, where we show as a function of the invariant mass the number of events per bin (upper) and the ratio $S$ of these events w.r.t. to the corresponding number of events in the SM (lower). We see that around the resonance peak  and for large invariant masses the deviation from the SM can be significant. It is, however, also evident that it will be complicated to see the threshold effect of the heavy top partners in the differential distribution. Of course the results depend on the chosen BP, bin size, luminosity, and final state, but they give a first rough idea of what might be expected in experimental conditions.

\begin{figure}[htb]
\centering
\includegraphics[width=0.6\textwidth]{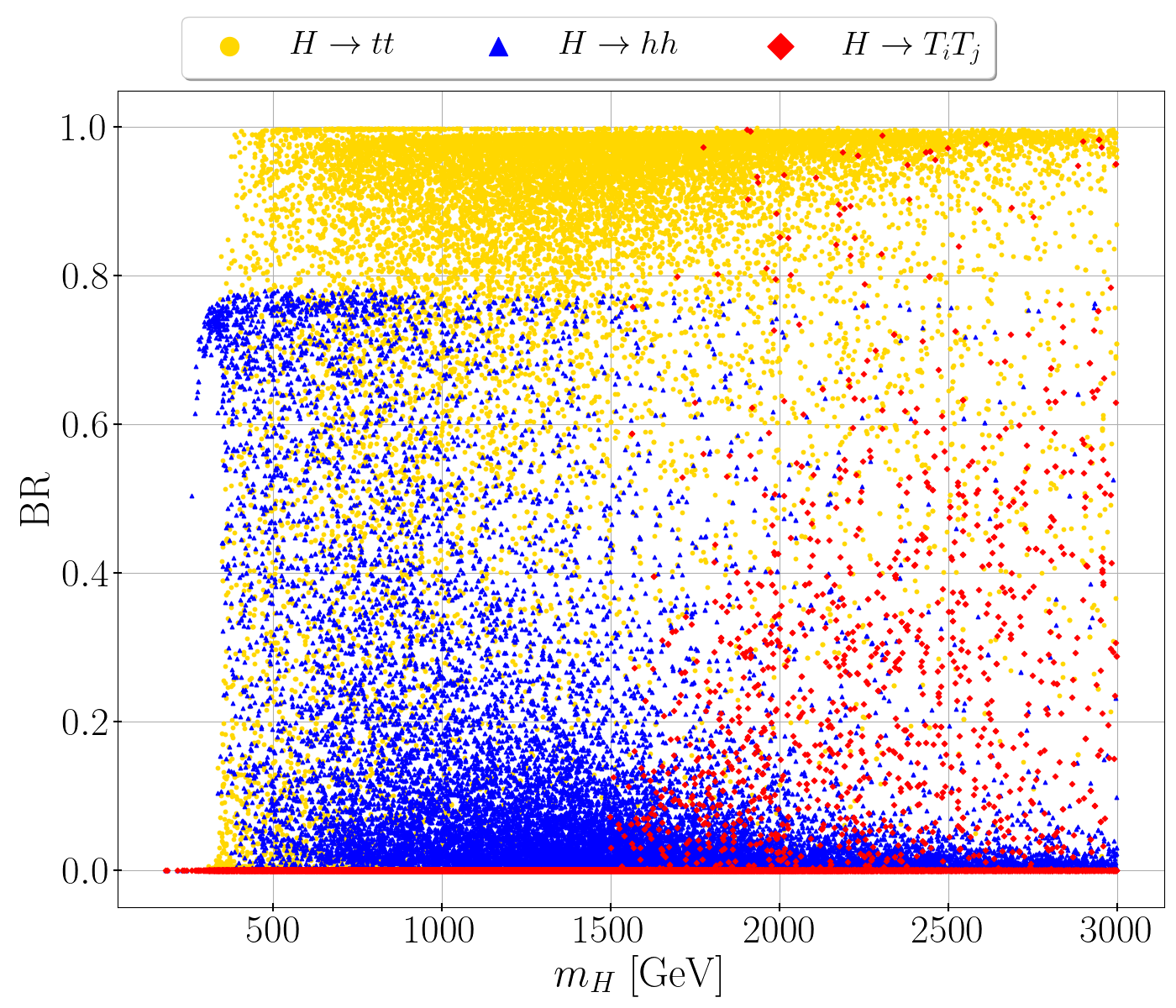} 
\caption{The BRs of the C2HDM heavy $H$ boson as a function of its mass in the following decay channels:  $tt$ (yellow), $hh$ (blue), and $T_iT_j$ with $i\ne j$ (red) and at least one being a heavy top quark with all possible final states been summed up.} 
\label{fig:BRH}
\end{figure}

\section{Comparison with the 2HDM Type-II, A2HDM and MSSM \label{sec:comparison}}
\label{sec:comparison2HDM}
\subsection{2HDM Type-II and A2HDM}

Since in the C2HDM we have aligned Yukawa couplings, it makes sense to compare our model with the elementary flavour-aligned 2HDM, the A2HDM~\cite{Pich:2009sp}. However, similar results are obtained for the four standard Yukawa types of the elementary 2HDM, so that we also illustrate below, e.g., the Type-II case. 

For the A2HDM, we have required that the Landau pole for the Yukawa couplings $\eta^F_k$ $(F=U,D,L\:,~k=1,2)$ does not appear at a certain energy scale $\Lambda$, using the following conditions,
\begin{align}
\label{eq:Llam}
|\eta^{F}_{k}(\Lambda)|^2 <\sqrt{4\pi}.
\end{align}
The Renormalisation Group (RG) running of these couplings is  evaluated with the 1-loop $\beta$ functions, which are derived by using {\tt SARAH}~\cite{Staub:2008uz}.
In solving the RG Equations (RGEs), we neglect the Yukawa coupling constants for the first and second generations of quarks and leptons. 
To alleviate the flavour constraints, we also impose the following relation for the Yukawa couplings in the mass basis,   ${\rho}_{b}/m_b={\rho}_{\tau}/m_\tau=0.1{\rho}_{t}/m_t$ (see Ref.~\cite{Davidson:2005cw} for the notation of the Yukawa couplings). 
We note that, after imposing the Landau pole condition,  Eq.~\eqref{eq:Llam}, $\rho_t$ is within the range of $-1.4\lesssim \rho_t v/(\sqrt{2}m_t)\lesssim 1.5$ for $\Lambda \gtrsim 5~{\rm TeV}$. 
The reason for this is that the top Yukawa couplings $\eta^F_k$ are enhanced by increasing $\rho_t$ and then they blow up by the RGE effects.

Apart from differences induced by the aforementioned additional $\xi$ dependence that  enters the C2HDM Yukawa and triple Higgs couplings (which was dealt with  extensively in Ref.~\cite{DeCurtis:2018zvh}), a noticeable feature is that the (composite) heavy CP-even Higgs resonance discussed in the previous section can be very wide, in comparison to the corresponding one of the A2HDM. The reason for this is twofold. On the one hand, the C2HDM is the low scale realisation of a strongly coupled theory, unlike the case of the A2HDM, which is a weakly coupled one, so that the actual strength of the relevant couplings entering our $gg\to hh$ process can  be larger in the former than the latter case. On the other hand, the presence in the C2HDM of heavy fermionic partners to the third generation SM quarks and leptons, unlike in the A2HDM, enables the heavy CP-even Higgs resonance to decay into new fermionic states: i.e., for the case of our 
analysis, $H\to T_9\bar T_i$ + c.c. ($i=1, ... 8$), with large partial decay widths. For illustration, we show in Fig.~\ref{fig:BRH} the main branching ratios of the heavy C2HDM Higgs boson $H$ as a function of its mass in the following decay channels: $tt$ (yellow), $hh$ (blue), and $T_iT_j$ with $i\ne j$ (red) and at least one being a heavy top quark with all possible final states been summed up. As can be inferred from the plot, for a heavy enough Higgs boson $H$, the decay into heavy top partners can be dominant.

\begin{figure}[htb]
\centering
\includegraphics[width=0.8\textwidth]{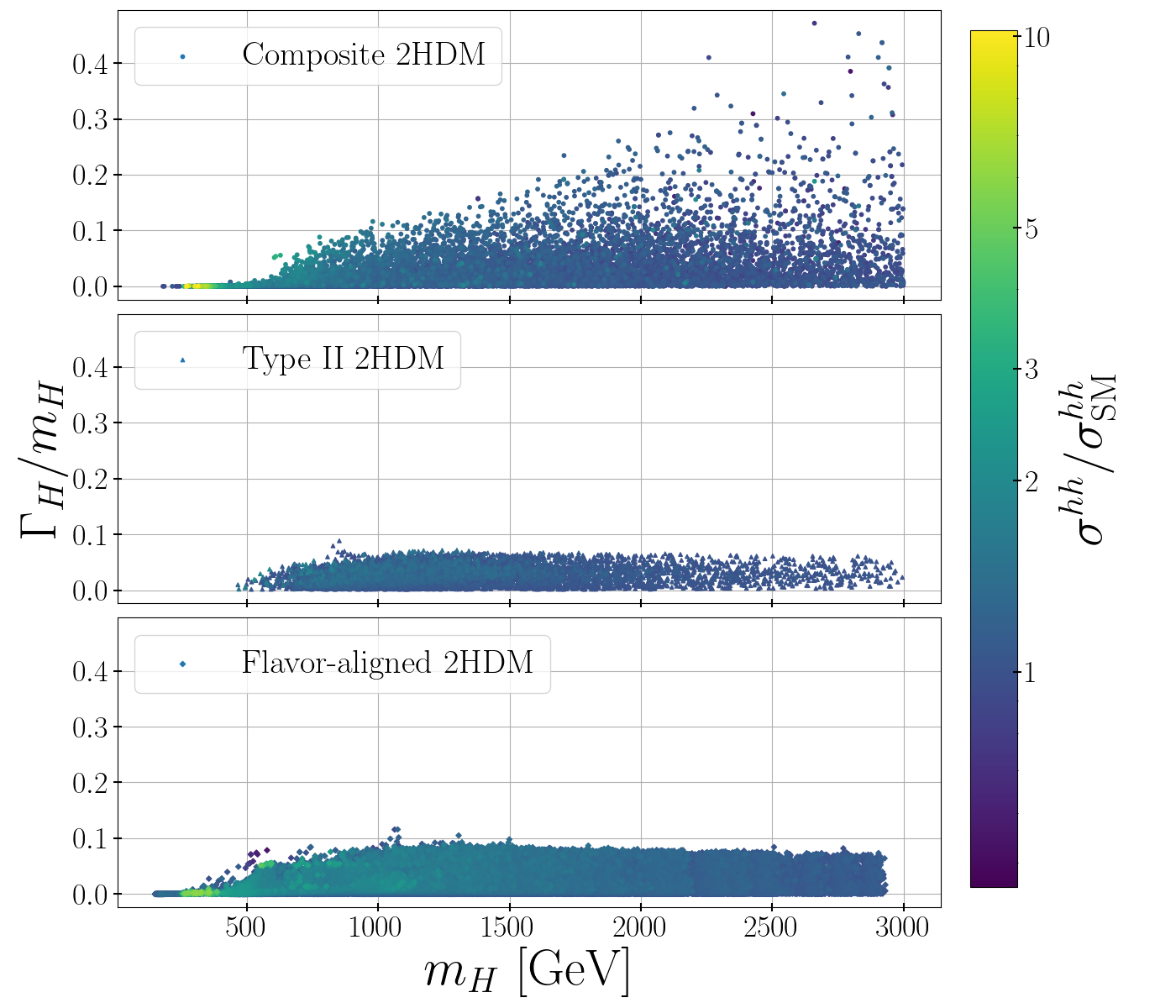} 
\caption{Total heavy Higgs decay width  normalised to its mass value as a function of the heavy Higgs mass for the allowed sample of points in the C2HDM (upper), 2HDM Type-II (middle) and A2HDM (lower). The colour code indicates the ratio of the di-Higgs cross section in the respective model to the SM value. \label{fig:GammaHmHsigtotComparisoncombined_constraints}}
\end{figure}

In Fig.~\ref{fig:GammaHmHsigtotComparisoncombined_constraints}, we show the ratio of the total width of the heavy Higgs boson and its mass value as a function of its mass for the C2HDM (upper), the 2HDM 
Type-II (middle),  
and the A2HDM (lower).\footnote{Note that compatibility with the flavour constraints forces the charged Higgs mass in the 2HDM type II to be rather heavy \cite{Haber:1999zh,Deschamps:2009rh,Mahmoudi:2009zx,Hermann:2012fc,Misiak:2015xwa,Misiak:2017bgg,Misiak:2020vlo}. In combination with the required compatibility with the electroweak precision data this excludes scalar mass values below about 480~GeV. As we mentioned at the end of section 2, this flavour bound can be easily relaxed in the C2HDM by requiring a small $\zeta_b$, which determines the $H^+ \bar t b$ coupling. The same requirement can be enforced for the A2HDM.} The colour code shows the ratio of the di-Higgs C2HDM cross section to the SM value. As can be inferred from the plot, the maximal width-to-mass ratio reaches $\sim 9\%$ in the 2HDM Type-II. In the flavour-aligned version A2HDM, it can go up to $\sim$ 12\%, while in the composite 2HDM it can become as large as $\sim$ 40\%. This is mainly due to the possible additional decays into final states with (at least one) heavy top partner. We also see that, for Higgs masses above about 1.5~TeV where the decay into heavy top partners is kinematically allowed, resonant production does not play a role.

In essence, in line with previously cited literature in the context of other collider processes \cite{DeCurtis:2016scv,DeCurtis:2016tsm,DeCurtis:2017gzi} (see also \cite{DeCurtis:2016gly}), 2HDM realisations from compositeness could be separated from elementary ones also with the help of SM-like 
di-Higgs production at the LHC.

\subsection{MSSM}
{We will defer the detailed comparison of the $gg\to hh$ between the C2HDM
and the MSSM, in the spirit of 
Ref. 
 \cite{DeCurtis:2018iqd}, to a forthcoming publication.
 However, based on the results of Ref.~\cite{Moretti:2023dlx} and those presented here, the most salient  phenomenological differences between the two theoretical paradigms, i.e., compositeness and supersymmetry, both  embedding a 2HDM Higgs sector, limited to the case of non-resonant $hh$ production\footnote{In fact, recall that  Ref.~\cite{Moretti:2023dlx} did not treat the resonant case via $gg\to H\to hh$.} are as follows.

 Firstly, $\lambda_{hhh}$ is more constrained in the MSSM than in the C2HDM, given that in supersymmetry the trilinear Higgs couplings are related to the gauge couplings, unlike in compositeness, so that the deviations seen here, driven by $\lambda_{hhh}$, are typically milder there. 
  Secondly, 
 in the MSSM (just like in the C2HDM) the $H$ can decay into top quark companions (i.e., the stops), but these decays are fewer in number ($H\to {\tilde t}_i{\tilde t}^*_j$ + c.c., $i,j=1, ... 2$) than in the compositeness case ($H\to T_i\bar T_j$ + c.c., $i,j=1, ... 9$) and with fewer degrees of freedoms (i.e., spin-0 vs spin-1/2 top quark companions, respectively), so that the contribution of these decays to $\Gamma_H$ is overall more significant in the C2HDM than in the MSSM, even when the ${\tilde t}_i$ masses can be smaller than the $T_i$ ones. Thirdly, it should be recalled that in the MSSM the so-called `alignment' and `decoupling' limits of  the Higgs sector are simultaneous whereas in the C2HDM the two can occur separately (see \cite{DeCurtis:2018iqd}), thereby implying that the pattern of interferences seen here between the topologies making up the amplitude for $gg\to hh$ in the C2HDM is necessarily different from that in the MSSM even in presence of a similar mass spectrum for the (lowest lying) top quark companions.  

Thus, echoing the message from the previous subsection, alongside other collider signatures \cite{DeCurtis:2018iqd},  
SM-like di-Higgs production at the LHC could offer the means to disentangle the C2HDM also from the MSSM.

\section{Conclusions \label{sec:summary}}

Both during Run 3 of the LHC and the HL-LHC phase, a target process will be di-Higgs ($hh$) production, as it is the experimental signature of the Higgs boson self-coupling and a sensitive probe of several BSM scenarios. In this paper, we performed a study of di-Higgs production via the dominant gluon-gluon fusion mode ($gg\to hh$), wherein  the relevant Feynman diagrams involving two SM-like Higgs bosons were computed through a sophisticated approach that enabled us to interpret possible signals of new physics in terms of squared Feynman amplitudes and their relative interferences, each of which has a well-defined mass and coupling structure. We illustrated the power of this procedure for the case of a specific realisation of compositeness, embedding a 2HDM sector as well as heavy quarks entering the loops of $gg\to hh$, called C2HDM. We have shown that the effects of either or both of the latter yield a change of the integrated cross section as well as peculiar kinematic features in its differential distributions with respect to the SM. These effects can in turn be traced back to the relevant diagrammatic and mass/coupling structures of the process, so that they  can eventually be used for diagnostic purposes.

In the case of non-resonant di-Higgs production, we have shown that sizable deviations from the SM cross section of tens of percent are possible at the inclusive level, in either direction, which then manifest themselves in exclusive spectra (chiefly, the $hh$ invariant mass) which are significantly different with respect to the SM ones. On the one hand, the threshold at $\approx 2m_t$ can be  modified in both position and height. On the other hand, another threshold emerges at $\approx 2 m_{T_i}$ (where $T_i$ is the lightest of the heavy top quarks), which clearly does not exist in the SM. In between these two values, we find that also the slope can be different between the SM and C2HDM. 

In the case of resonant di-Higgs production, wherein a $s$-channel diagram with $H$ propagation enters, whenever
$m_H> 2m_h$, we find that the typical Breit-Wigner shape, determined primarily by $\Gamma_H$, is subject to significant distortions by interferences with other topologies, both SM ones and C2HDM specific ones (i.e., involving heavy top quarks). Furthermore, the values of $\Gamma_H/m_H$ found in the C2HDM, of up to ${\cal O}(40-50\%)$, are generally larger than those found in the elementary 2HDM with a similar Yukawa structure, i.e., an aligned one (A2HDM), of at most {${\cal O}(18\%)$}, owing to both additional contributions of composite origin to the $Ht\bar t$ coupling and the fact that $H\to t\bar T_i$ (+ c.c.) decays can onset.

This altogether opens the prospect of using di-Higgs production at the LHC as a proxy to this specific kind of BSM physics, which is notably different from not only the corresponding {elementary 2HDM} structures but also from the 2HDM counterpart of another viable theory of the EW scale, i.e., supersymmetry, in the form of the MSSM (as can be deduced from a comparison with a similar calculation existing in literature in this scenario). Indeed, in order to aid the experimental pursuit of the C2HDM effects described here, we have already released in this paper several BPs representative of parameter space configurations enabling both non-resonant and resonant SM-like di-Higgs production and decay plus we will eventually make the computing tools developed here also available in the public domain.

\bigskip
\subsubsection*{Acknowledgments}
The work of FE and MM is supported by the BMBF-Project 05H21VKCCA. 
SM is supported in part through the
NExT Institute, the STFC Consolidated Grant No. ST/L000296/1 and the 
Knut and Alice Wallenberg foundation under the
grant KAW 2017.0100.
KS is supported by JSPS KAKENHI Grant No. 20H01894, No. 23KJ0086, and 
the National Science Centre, Poland, under research Grant No. 2020/38/E/ST2/00243. 
This work is partially supported by ICSC – Centro Nazionale di Ricerca in High Performance Computing, Big Data and Quantum Computing, funded by the European Union under the project  NextGenerationEU.

\newpage
\appendix
\section*{Appendix}
\section{Analytic Expressions for the LO Cross Section}
\label{sec:appendix1}
\subsection{Notation}
In the derivation of the analytic expressions, we followed the conventions applied in \cite{Plehn:1996wb,Grober:2010yv,Gillioz:2012se,Grober:2015cwa,Grober:2016wmf}.
We take all momenta as ingoing, i.e.,~the Mandelstam variables read
\begin{align}\label{eq:mandelstam}
\hat{s}=(p_1 + p_2)^2, \; \; \; \hat{t}=(p_1 + p_3)^2, \; \;\; \hat{u}=(p_2 + p_3)^2,
\end{align}
where $p_1$ and $p_2$ are the gluon and $p_3$ and $p_4$ are the Higgs momenta.
Furthermore, we define the following projectors
\begin{align}
A_1^{\mu \nu}&=g^{\mu \nu} - \frac{p_1^{\nu}p_2^{\mu}}{(p_1 \cdot p_2)}\,, \\
A_2^{\mu \nu}&=g^{\mu \nu} + \frac{p_3^{2}p_1^{\nu}p_2^{\mu}}{p_{T}^{2}(p_1 \cdot p_2)} - \frac{2(p_3 \cdot p_2)p_1^{\nu}p_3^{\mu}}{p_{T}^{2}(p_1 \cdot p_2)} - \frac{2(p_3 \cdot p_1) p_3^{\nu} p_2^{\mu}}{p_{T}^{2}(p_1 \cdot p_2)} + \frac{2 p_3^{\mu} p_3^{\nu}}{p_T^{2}}\,,\\
p_T^{2}&= 2 \frac{(p_1 \cdot p_3)(p_2 \cdot p_3)}{(p_1 \cdot p_2)} - p_3^2\,, \label{eq:pTspira}
\end{align}
with
\begin{align}
A_1 \cdot A_2 =0, \; \; \; \;  A_1 \cdot A_1 =A_2 \cdot A_2 =2 \,.
\end{align}
The total amplitude can be decomposed in the amplitudes for the triangle and the box diagrams,
\begin{align}
\mathcal{A}(gg \to hh) = \mathcal{A}_\triangle + \mathcal{A}_\Box \,,
\label{eq:amplitude}
\end{align}
with the amplitude for the triangle diagrams  given by 
\begin{align}
\mathcal{A}_{\triangle}=\frac{\alpha_sG_{F} \sqrt{2}}{4\pi} \hat{s} A_1^{\mu \nu} \epsilon_{\mu}^{a}\epsilon_\nu^{b} \delta_{ab} \sum_{i=1}^{9}C_{i,\triangle}^{hh}F_\triangle (m_i)\,,
\end{align}
and the one for the box diagrams by
\begin{align}
\mathcal{A}_{\Box}=\frac{\alpha_sG_{F} \sqrt{2}}{4\pi} \hat{s} \epsilon_{\mu}^{a}\epsilon_\nu^{b} \delta_{ab} \sum_{i=1}^{9}\sum_{j=1}^{9}
& \left[A_1^{\mu \nu} \left(C_{i,j,\Box}^{hh} F_\Box (m_i,m_j) + C_{i,j,\Box,5}^{hh} F_{\Box,5} (m_i,m_j)  \right) \right. \nonumber \\
+ & \left .A_2^{\mu \nu} \left(C_{i,j,\Box}^{hh} G_\Box (m_i,m_j) + C_{i,j,\Box,5}^{hh} G_{\Box,5} (m_i,m_j)  \right) \right].
\end{align}
The generalised couplings $C^{hh}_{i,\triangle}$, $C^{hh}_{ij,\Box}$, and $C^{hh}_{ij,\Box,5}$ have been defined in Eqs.~(\ref{eq:CFactors1}) to (\ref{eq:CFactors3}), and the form factors $F_\triangle$, $F_{\Box}$, $F_{\Box,5}$, $G_{\Box}$, and $G_{\Box,5}$ are given in App.~\ref{app:formfactors}. The differential cross section w.r.t.~$\hat{t}$, derived from the amplitude Eq.~(\ref{eq:amplitude}), has been given in Eq. \eqref{eq:diffcxn}. To obtain the full partonic cross section we have to integrate it, i.e.,
\begin{align}
\hat{\sigma}(gg \to hh)&= \int_{\hat{t_{-}}}^{\hat{t_{+}}} d\hat{t} \, \frac{\mathrm{d} \hat{\sigma}(gg \to hh)}{\mathrm{d}\hat{t}}, \; \; \; 
\hat{t}_{\pm}=\frac{-\hat{s}}{2} \left( 1- 2\frac{m_h^2}{\hat{s}} \mp \sqrt{1- \frac{4 m_h^2}{\hat{s}}} \right).
\end{align}
The full hadronic cross section is obtained by folding the partonic cross section with the gluon luminosity $\frac{\mathrm{d}\mathcal{L}^{gg}}{\mathrm{d}\tau}$ and integrate over the c.m. energy, i.e.,
\begin{align}
\sigma(pp\to gg \to hh)= \int_{\tau_0}^{1} \mathrm{d}\tau \frac{\mathrm{d}\mathcal{L}^{gg}}{\mathrm{d}\tau}\hat{\sigma}(\hat{s}=\tau s)\,, \; \; \;  \tau_0 = \frac{4m_h^2}{s}.
\end{align}

\subsection{Form Factors}
\label{app:formfactors}
For the form factors, we obtain
\beq
F_\triangle (m_i) &=& \frac{2m_i}{\hat{s}} [2+(4 m_i^2-\hat{s})
C_{12}], \\
F^{hh}_\Box (m_i,m_j) &=& \frac{1}{\hat{s}} \left[4+8m_i^2C_{12} +
\frac{2}{\hat{s}}[m_h^2-(m_i+m_j)^2](\hat{u}-m_{h}^2) (C_{23}+C_{14})
\right. \nonumber \\
&& + \frac{2}{\hat{s}}[m_h^2-(m_i+m_j)^2](\hat{t}-m_{h}^2) (C_{13}+C_{24}) 
\nonumber \\
&& 
+ 2[(m_i+m_j)(2m_i^2 (m_i+m_j)-m_i \hat{s})-m_i^2 (\hat{u}+\hat{t})]
(D_{123}+D_{213}+D_{132}) \nonumber \\
&& \left. -\frac{2}{\hat{s}} [\hat{t}
  \hat{u}-m_h^4+s(m_j^2-m_i^2)][m_h^2-(m_i+m_j)^2] D_{132} \right.], \\
%
G^{hh}_\Box (m_i,m_j) &=& \frac{1}{\hs (\hu \htt -m_h^4)} \left[ (\htt^2
  +\hu^2 - 4 (m_j^2+m_i m_j) (\htt+\hu) + 4(m_j-m_i)(m_i+m_j)^3
   \right. \nonumber \\
&& +2m_h^4) \hs C_{12}+ (m_h^4+\htt^2-2\htt (m_i+m_j)^2) ((C_{13}+C_{24}) (\htt-m_h^2)-\hs 
\htt D_{213}) \nonumber \\
&& + (m_h^4+\hu^2-2\hu (m_i+m_j)^2) ((C_{23}+C_{14}) (\hu-m_h^2)-\hs 
\hu D_{123}) \nonumber \\
&& - (\htt^2+\hu^2-2m_h^4) (\htt+\hu-2(m_i+m_j)^2) C_{34} \nonumber \\
&& - (\htt+\hu-2(m_i+m_j)^2)((\htt \hu-m_h^4)(m_i^2+m_j^2)+\hs
(m_i^2-m_j^2)^2) \nonumber \\
&& (D_{123}+D_{213}+D_{132}) +2(m_i^2-m_j^2)[-(\hs+m_h^2)(m_h^4+\hu^2
\nonumber \\
&& -2\hu(m_i+m_j)^2)+(-\hs\hu-\hu^2-m_h^4)((m_i+m_j)^2-m_h^2) \nonumber \\
&&+m_h^2(\hu-m_h^2)^2]D_{123}+2(m_i^2-m_j^2)[-(\hs+m_h^2)(m_h^4+\htt^2-2\htt (m_i+m_j)^2)\nonumber \\
&&\left. +
(-\hs\htt-\htt^2-m_h^4)((m_i+m_j)^2-m_h^2)+m_h^2(\htt-m_h^2)^2]D_{213} \right].
\eeq
The form factors $F^{hh}_{\Box,5} (m_i,m_j)$ and $G^{hh}_{\Box,5} (m_i,m_j)$ are obtained via
\beq
F^{hh}_{\Box,5} (m_i,m_j) &=& - F^{hh}_\Box (m_i,-m_j), \\
G^{hh}_{\Box,5} (m_i,m_j) &=& - G^{hh}_\Box (m_i,-m_j) \;.
\eeq
The $C_{ij}$ and $D_{ijk}$ integrals in the expressions are defined as 
\beq
C_{ij}(m_1^2,m_2^2,m_3^2) &=& \int \frac{d^4 q}{i\pi^2}
\frac{1}{(q^2-m_1^2)((q+p_i)^2-m_2^2)((q+p_i+p_j)^2-m_3^2)},   \\ 
D_{ijk}(m_1^2,m_2^2,m_3^2,m_4^2) &=& \nonumber\\
&& \hspace*{-2cm}\int \frac{d^4 q}{i\pi^2}
\frac{1}{(q^2-m_1^2)((q+p_i)^2-m_2^2)((q+p_i+p_j)^2-m_3^2)((q+p_i+p_j+p_k)^2-m_4^2)},  \nonumber\\
\eeq
and we use the abbreviations
\beq
C_{12} &\equiv& C_{12} (m_i^2,m_i^2,m_i^2), \nonumber \\
C_{13} &\equiv& C_{13} (m_i^2,m_i^2,m_j^2), \nonumber \\
C_{14} &\equiv& C_{14} (m_i^2,m_i^2,m_j^2), \nonumber \\
C_{23} &\equiv& C_{23} (m_i^2,m_i^2,m_j^2), \nonumber \\
C_{24} &\equiv& C_{24} (m_i^2,m_i^2,m_j^2), \nonumber \\
C_{34} &\equiv& C_{34} (m_i^2,m_j^2,m_i^2), \nonumber \\
D_{123} &\equiv& D_{123} (m_i^2,m_i^2,m_i^2,m_j^2), \nonumber \\
D_{213} &\equiv& D_{213} (m_i^2,m_i^2,m_i^2,m_j^2), \nonumber \\
D_{132} &\equiv& D_{132} (m_i^2,m_i^2,m_j^2,m_j^2) \;.
\eeq
\section{Derivation of the $p_T$ Distribution}
\label{app:pT}
In the following, we use the same conventions as in App. \ref{sec:appendix1} for the Mandelstam variables and the momenta, but now allow for different final state particles, i.e., masses $m_3$ and $m_4$, $m_3 \ne m_4$. We start with the formula for the full hadronic cross section
\begin{align}
\sigma_{pp\to \phi \phi^\prime} &= \int_{\tau_0}^1 \mathrm{d}\tau \, \frac{\mathrm{d} {\mathcal{L}}^{gg}}{\mathrm{d}\tau} \, {\hat{\sigma}}_{gg\to \phi \phi^\prime} (\hat{s} = \tau s) \notag \\
&=\int_{\tau_0}^1 \mathrm{d}\tau \int^{t_+}_{t_-}\mathrm{d}\hat{t}
\frac{\mathrm{d} {\mathcal{L}}^{gg}}{\mathrm{d}\tau} \left(\frac{\mathrm{d} \hat{\sigma}_{gg\to \phi \phi^\prime}}{\mathrm{d} \hat{t}}\right)_{\hat{s}=\tau s} \;,
\end{align}
with  
\begin{align}
\tau_0&=\frac{(m_3+m_4)^2}{s} \;, \\
\hat{t}_{\pm}&=\frac{1}{2} \left( m_3^2+m_4^2 - \hat{s} \pm \lambda^{\frac{1}{2}}(\hat{s},m_3^2,m_4^2) \right),
\end{align}
where we used the Källén function
\begin{align}
\lambda(a,b,c)\equiv a^2+b^2+c^2-2ab-2ac-2bc.
\end{align}
We obtain the invariant mass distribution by omitting the integration over $\tau$. Now we want to derive the $p_T$ distribution. First we have to change the integration variable from $\hat{t}$ to $p_T^2$ and then change the integration order. We have
\begin{align}
\hat{s}=4 E_1^2 \quad \Rightarrow E_1=\frac{\sqrt{\hat{s}}}{2}= |\vec{p}_1| \,,
\end{align}
and obtain
\begin{align}
\hat{t}=p_1^2 + p_3^2 +2p_1p_3=m_3^2 -2E_1E_3 +2  |\vec{p}_1|  |\vec{p}_3| \cos \theta,
\end{align}
where $\theta$ is the angle between $\vec{p}_1$ and $\vec{p}_3$. 
With this we can express $\cos\theta$ as
\begin{align}\label{eq:costheta}
\cos \theta = \frac{\hat{t}-m_3^2 +2E_1E_3}{2 |\vec{p}_1|  |\vec{p}_3|}.
\end{align}
To obtain the transverse momentum, we have to project $\vec{p}_3$ to the transverse plain, i.e.,
\begin{align}\label{eq:pTprojection}
p_T^2=\sin\theta^2 |\vec{p}_3|^3 = (1- \cos\theta^2)|\vec{p}_3|^2.
\end{align} 
Finally, we also need the absolute value of $\vec{p}_3$ which can be obtained from 
\begin{align}
\hat{s}&=(p_3+p_4)^2=(E_3+E_4)^2=(\sqrt{m_3^2+|\vec{p}|^2}+\sqrt{m_4^2+|\vec{p}|^2})^2 \;, \\ &\Rightarrow |\vec{p}|^2 = \frac{1}{4\hat{s}}\lambda(\hat{s},m_3^2,m_4^2). \nonumber
\end{align}
Now using $|\vec{p}_1|=E_1$ and $|\vec{p}_3|=|\vec{p}|$ and combining Eqs. \eqref{eq:costheta} and \eqref{eq:pTprojection}, we obtain
\begin{align}
p_T^2=\frac{(m_4^2-\hat{s}-\hat{t})\hat{t}+m_3^2(\hat{t}-m_4^2)}{\hat{s}} \;,
\end{align}
which coincides with the definition in Eq. \eqref{eq:pTspira}. Solving this equation for $\hat{t}$ results in
\begin{align}
\hat{t}_{1,2}=\frac{1}{2} \left( m_3^2+m_4^2 -\hat{s} \pm \sqrt{\lambda(\hat{s},m_3^2,m_4^2)-4p_T^2\hat{s}} \right) .
\end{align}
When solving this equation we have to take the square root of $(\hat{t} + \frac{\hat{s}-m_3^2-m_4^2}{2})^2$. So we have to split up the $\hat{t}$ integration from $t_{-}$ to $t_m$ and from $t_m$ to $t_{+}$, i.e., the parts where the expression is positive or negative, with $t_m$ being the point where this bracket vanishes,
\begin{align}
\hat{t}_m&=\frac{m_3^2+m_4^2-\hat{s}}{2}.
\end{align}
It holds 
\begin{align}
p_T^2(\hat{t}_{\pm})&=0, \\
\frac{\mathrm{d}p_T^2}{\mathrm{d}\hat{t}}&=\frac{m_3^2+m_4^2-\hat{s}-2\hat{t}}{\hat{s}}  \stackrel{!}{=}0 \quad \Rightarrow \hat{t}_{\text{max}}=\hat{t}_m \\
\frac{\mathrm{d}^2p_T^2}{\mathrm{d}\hat{t}^2}&=-\frac{2}{\hat{s}}<0 \rightarrow \text{Maximum} \\
p_T^2(\hat{t}_{m})&=\frac{\lambda(\hat{s},m_3^2,m_4^2)}{4\hat{s}} \equiv p_{T,\text{max}}^2.
\end{align}
Thus the transverse momentum vanishes at the integration boundaries of $\hat{t}$ and reaches it maximum at $\hat{t}_m$. We then have to integrate from $0$ to $p_{T,\text{max}}^2$.

We also need the substitution rule,
\begin{align}
\mathrm{d}\hat{t}=\mathrm{d}p_T^2\frac{\hat{s}}{m_3^2+m_4^2-\hat{s} -2 \hat{t}}.
\end{align}
Thus, we change the $\hat{t}$ integration to
\begin{align}
\int_{\hat{t}_{-}}^{\hat{t}_{+}}\mathrm{d}\hat{t} \;\frac{\mathrm{d} \hat{\sigma}}{\mathrm{d} \hat{t}} &=\int_{\hat{t}_{-}}^{\hat{t}_{m}}\mathrm{d}\hat{t}\;\frac{\mathrm{d} \hat{\sigma}}{\mathrm{d} \hat{t}} + \int_{\hat{t}_{m}}^{\hat{t}_{+}}\mathrm{d}\hat{t} \;\frac{\mathrm{d} \hat{\sigma}}{\mathrm{d} \hat{t}}  \\
 &\hspace*{-1cm} = \int_{0}^{p_{T,\text{max}}^2}\left( \mathrm{d}p_T^2 \frac{\hat{s}}{m_3^2+m_4^2-\hat{s} -2 \hat{t}} \;\frac{\mathrm{d} \hat{\sigma}}{\mathrm{d} \hat{t}} \right)_{\hat{t}=\hat{t}_2}
+\int_{p_{T,\text{max}}^2}^{0}\left(\mathrm{d}p_T^2\frac{\hat{s}}{m_3^2+m_4^2-\hat{s} -2 \hat{t}}\;\frac{\mathrm{d} \hat{\sigma}}{\mathrm{d} \hat{t}} \right)_{\hat{t}=\hat{t}_1}  \nonumber \\
 &\hspace*{-1cm} =\int_{0}^{p_{T,\text{max}}^2}\mathrm{d}p_T^2 \left[ \left(\frac{\hat{s} }{m_3^2+m_4^2-\hat{s} -2 \hat{t}}\;\frac{\mathrm{d} \hat{\sigma}}{\mathrm{d} \hat{t}}\right)_{\hat{t}=\hat{t}_2(p_T^2)} -\left( \frac{\hat{s}}{m_3^2+m_4^2-\hat{s} -2 \hat{t}}\;\frac{\mathrm{d} \hat{\sigma}}{\mathrm{d} \hat{t}}\right)_{\hat{t}=\hat{t}_1(p_T^2)} \right].
\end{align}
We have $p_T^2 \in [0,p_{T,\text{max}}^2]$ and $\tau \in [\tau_0,1]$.
To change the integration boundaries, we need to solve the inequality
\begin{align}
p_T^2 \leq p_{T,\text{max}}^2 = \frac{\lambda(\tau s,m_3^2,m_4^2)}{4\tau s}
\end{align}
for $\tau$. After some simplification this leads to the relation
\begin{align}
4(p_T^2+m_3^2)(p_T^2+m_4^2)\leq (\tau s -2 p_T^2 -m_3^2 -m_4^2)^2.
\end{align}
The question is whether the term in brackets on the right-hand side is always positive. This is the case, since
\begin{align}
\tau s -2 p_T^2 -m_3^2 -m_4^2 \geq \tau s -2 p_{T,\text{max}}^2 -m_3^2 -m_4^2 &= \frac{\tau^2 s^2 - (m_3^2+m_4^2)^2}{2\tau s} \\
 \geq  \frac{\tau_0^2 s^2 - (m_3^2+m_4^2)^2}{2\tau s}&= \frac{4m_3m_4(m_3^2+m_4^2)}{\tau s} \geq 0.
\end{align}
So the square root can easily be taken, and we obtain the inequality
\begin{align}
\tau \geq \frac{1}{s} \left( 2\sqrt{(p_T^2+m_3^2)(p_T^2+m_4^2)} +2p_T^2 +m_3^2 +m_4^2 \right) \equiv \tau_{\text{min}}.
\end{align}
Thus, we have the new integration bounds
\begin{align}
p_T^2 \in [0, \frac{\lambda(s,m_3^2,m_4^2)}{4s}], \quad \tau \in [\frac{1}{s} \left( 2\sqrt{(p_T^2+m_3^2)(p_T^2+m_4^2)} +2p_T^2 +m_3^2 +m_4^2 \right),1],
\end{align}
and the full hadronic cross section can be expressed as
\begin{align}
\sigma_{pp\to \phi \phi^\prime}  =
     \int_0^{p_{T,\text{max}}^2} \mathrm{d}p_T^2
\left(\frac{\mathrm{d}\hat{\sigma}}{\mathrm{d}p_T^2}\right),
\end{align}
where
\begin{align}
\frac{\mathrm{d}\hat{\sigma}}{\mathrm{d}p_T^2}  &=
     \int_{\tau_{\rm min}}^1 d\tau
    \frac{d {\cal L}^{gg}}{d\tau}  \notag \\
&\times
    \left[ \left(\frac{\hat{s} }{m_3^2+m_4^2-\hat{s} -2 \hat{t}}\;\frac{\mathrm{d} \hat{\sigma}}{\mathrm{d} \hat{t}}\right)_{\hat{t}=\hat{t}_2(p_T^2)} -\left( \frac{\hat{s}}{m_3^2+m_4^2-\hat{s} -2 \hat{t}}\;\frac{\mathrm{d} \hat{\sigma}}{\mathrm{d} \hat{t}}\right)_{\hat{t}=\hat{t}_1(p_T^2)} \right]. 
\end{align}

\section{BPs \label{app:bps}}
In Tab.~\ref{tab:bps}, we summarise the input values of the BPs discussed in the numerical analysis.
\begin{table}[]
    \centering
    \scalebox{0.7}{
    \begin{tabular}{c|cccccc}
       BP & $f$ [GeV] & $\Delta_L$ [GeV] & $\Delta_R$ [GeV] & $Y_1$ [GeV] & $Y_2$ [GeV] & $g_\rho$\\
       \hline 
    BP 1  &   1139.21 &  $\left(\begin{array}{c} 649.392  \\ -1787.9 \end{array}\right)$ & $\left(\begin{array}{c} -7244.85 \\ 4633.51 \end{array}\right)$  &
$\left(\begin{array}{cc} -406.903 & 421.383 \\ -910.863 & -1651.99 \end{array}\right)$ & $\left(\begin{array}{cc} 3996.82 & 2846.41 \\ 2265.86 & 518.944 \end{array}\right)$ & 7.02515 \\
BP 2 & 821.74 & $\left(\begin{array}{c} 5172.74 \\ -3835.24 \end{array}\right)$ & $\left(\begin{array}{c} -2850.8 \\ -759.562 \end{array}\right)$ & 
$\left(\begin{array}{cc} 3194.11 & 2467.64 \\ 2748.76 & 1489.54 \end{array}\right)$ & $\left(\begin{array}{cc} 457.272 & -1135.19 \\ 5946.7 & -3126.3 \end{array}\right)$ & 7.87477 \\
BP 3 & 795.639 & $\left(\begin{array}{c} -168.309 \\  1137.24 \end{array}\right)$ & $\left(\begin{array}{c} -2548.98 \\ -2181.22 \end{array}\right)$ &  $\left(\begin{array}{cc} -1808.81 & -695.861 \\ 3507.5 & -320.533 \end{array}\right)$ &  $\left(\begin{array}{cc} 4348.75 & 399.558 \\ -4182.72 & -1915.42 \end{array}\right)$ & 6.7523 \\
BP 4 & 750.293 & $\left(\begin{array}{c} -1007.88 \\ -1351.26 \end{array}\right)$ & $\left(\begin{array}{c} 1844.02 \\  1713.76 \end{array}\right)$ & $\left(\begin{array}{cc} 709.119 & -884.948 \\ -5689.43 & 3420.92 \end{array}\right)$ & $\left(\begin{array}{cc} 2833.62 & -2811.59 \\ 
5092.76 & 3134.5 \end{array}\right)$ & 8.6289 \\
\end{tabular}
    }
    \caption{Input values of the BPs analysed in the paper}
\label{tab:bps}
\end{table}

\newpage

\bibliography{Composite2HDMHH}

\providecommand{\href}[2]{#2}\begingroup\raggedright\begin{thebibliography}{100}

\bibitem{ATLAS:2022vkf}
{\scshape ATLAS} collaboration, \emph{{A detailed map of Higgs boson
  interactions by the ATLAS experiment ten years after the discovery}},
  \href{http://dx.doi.org/10.1038/s41586-022-04893-w}{\emph{Nature} \textbf{
  607} (2022) 52--59}, [\href{https://arxiv.org/abs/2207.00092}{{\texttt
  2207.00092}}].

\bibitem{CMS:2022dwd}
{\scshape CMS} collaboration, A.~Tumasyan et~al., \emph{{A portrait of the
  Higgs boson by the CMS experiment ten years after the discovery}},
  \href{http://dx.doi.org/10.1038/s41586-022-04892-x}{\emph{Nature} \textbf{
  607} (2022) 60--68}, [\href{https://arxiv.org/abs/2207.00043}{{\texttt
  2207.00043}}].

\bibitem{Higgs:1964pj}
P.~W. Higgs, \emph{{Broken Symmetries and the Masses of Gauge Bosons}},
  \href{http://dx.doi.org/10.1103/PhysRevLett.13.508}{\emph{Phys. Rev. Lett.}
  \textbf{ 13} (1964) 508--509}.

\bibitem{Englert:1964et}
F.~Englert and R.~Brout, \emph{{Broken Symmetry and the Mass of Gauge Vector
  Mesons}}, \href{http://dx.doi.org/10.1103/PhysRevLett.13.321}{\emph{Phys.
  Rev. Lett.} \textbf{ 13} (1964) 321--323}.

\bibitem{Guralnik:1964eu}
G.~S. Guralnik, C.~R. Hagen and T.~W.~B. Kibble, \emph{{Global Conservation
  Laws and Massless Particles}},
  \href{http://dx.doi.org/10.1103/PhysRevLett.13.585}{\emph{Phys. Rev. Lett.}
  \textbf{ 13} (1964) 585--587}.

\bibitem{Kibble:1967sv}
T.~W.~B. Kibble, \emph{{Symmetry breaking in nonAbelian gauge theories}},
  \href{http://dx.doi.org/10.1103/PhysRev.155.1554}{\emph{Phys. Rev.} \textbf{
  155} (1967) 1554--1561}.

\bibitem{Djouadi:1999gv}
A.~Djouadi, W.~Kilian, M.~M{\"u}hlleitner and P.~M. Zerwas, \emph{{ Testing
  Higgs self-couplings at $e^+e^-$ linear colliders}},
  \href{http://dx.doi.org/10.1007/s100529900082}{\emph{Eur. Phys. J.} \textbf{
  C10} (1999) 27--43}, [\href{https://arxiv.org/abs/hep-ph/9903229}{{\texttt
  hep-ph/9903229}}].

\bibitem{Djouadi:1999rca}
A.~Djouadi, W.~Kilian, M.~M{\"u}hlleitner and P.~M. Zerwas, \emph{{Production
  of neutral Higgs boson pairs at LHC}},
  \href{http://dx.doi.org/10.1007/s100529900083}{\emph{Eur. Phys. J.} \textbf{
  C10} (1999) 45--49}, [\href{https://arxiv.org/abs/hep-ph/9904287}{{\texttt
  hep-ph/9904287}}].

\bibitem{atlaspaperdihiggs}
A.~Collaboration, \emph{{Constraining the Higgs boson self-coupling from
  single- and double-Higgs production with the ATLAS detector using pp
  collisions at $\sqrt{s}=13$ TeV}}, {\emph{ATLAS-CONF-2022-50} (2022) }.

\bibitem{CMS:2022hgz}
C.~Collaboration, \emph{{Search for nonresonant Higgs boson pair production in
  final state with two bottom quarks and two tau leptons in proton-proton
  collisions at $\sqrt{s}$ = 13 TeV}},
  \href{https://arxiv.org/abs/2206.09401}{{\texttt 2206.09401}}.

\bibitem{Baglio:2012np}
J.~Baglio, A.~Djouadi, R.~Grober, M.~M. M{\"u}hlleitner, J.~Quevillon and
  M.~Spira, \emph{{The measurement of the Higgs self-coupling at the LHC:
  theoretical status}},
  \href{http://dx.doi.org/10.1007/JHEP04(2013)151}{\emph{JHEP} \textbf{ 04}
  (2013) 151}, [\href{https://arxiv.org/abs/1212.5581}{{\texttt 1212.5581}}].

\bibitem{deFlorian:2016spz}
{\scshape LHC Higgs Cross Section Working Group} collaboration, D.~de~Florian
  et~al., \emph{{Handbook of LHC Higgs Cross Sections: 4. Deciphering the
  Nature of the Higgs Sector}},
  \href{https://arxiv.org/abs/1610.07922}{{\texttt 1610.07922}}.

\bibitem{DiMicco:2019ngk}
J.~Alison et~al., \emph{{Higgs boson potential at colliders: Status and
  perspectives}},
  \href{http://dx.doi.org/10.1016/j.revip.2020.100045}{\emph{Rev. Phys.}
  \textbf{ 5} (2020) 100045}, [\href{https://arxiv.org/abs/1910.00012}{{\texttt
  1910.00012}}].

\bibitem{Glover:1987nx}
E.~W.~N. Glover and J.~J. van~der Bij, \emph{{Higgs boson pair production via
  gluon fusion}},
  \href{http://dx.doi.org/10.1016/0550-3213(88)90083-1}{\emph{Nucl. Phys.}
  \textbf{ B309} (1988) 282--294}.

\bibitem{Dicus:1987ic}
D.~A. Dicus, C.~Kao and S.~S.~D. Willenbrock, \emph{{Higgs Boson Pair
  Production From Gluon Fusion}},
  \href{http://dx.doi.org/10.1016/0370-2693(88)90202-X}{\emph{Phys. Lett. B}
  \textbf{ 203} (1988) 457--461}.

\bibitem{Plehn:1996wb}
T.~Plehn, M.~Spira and P.~M. Zerwas, \emph{{Pair production of neutral Higgs
  particles in gluon-gluon collisions}},
  \href{http://dx.doi.org/10.1016/0550-3213(96)00418-X}{\emph{Nucl. Phys. B}
  \textbf{ 479} (1996) 46--64},
  [\href{https://arxiv.org/abs/hep-ph/9603205}{{\texttt hep-ph/9603205}}].

\bibitem{Grazzini:2018bsd}
M.~Grazzini, G.~Heinrich, S.~Jones, S.~Kallweit, M.~Kerner, J.~M. Lindert
  et~al., \emph{{Higgs boson pair production at NNLO with top quark mass
  effects}}, \href{http://dx.doi.org/10.1007/JHEP05(2018)059}{\emph{JHEP}
  \textbf{ 05} (2018) 059}, [\href{https://arxiv.org/abs/1803.02463}{{\texttt
  1803.02463}}].

\bibitem{Gianotti:2002xx}
F.~Gianotti et~al., \emph{{Physics potential and experimental challenges of the
  LHC luminosity upgrade}},
  \href{http://dx.doi.org/10.1140/epjc/s2004-02061-6}{\emph{Eur. Phys. J. C}
  \textbf{ 39} (2005) 293--333},
  [\href{https://arxiv.org/abs/hep-ph/0204087}{{\texttt hep-ph/0204087}}].

\bibitem{Cepeda:2019klc}
M.~Cepeda et~al., \emph{{Report from Working Group 2}: {Higgs Physics at the
  HL-LHC and HE-LHC}},
  \href{http://dx.doi.org/10.23731/CYRM-2019-007.221}{\emph{CERN Yellow Rep.
  Monogr.} \textbf{ 7} (2019) 221--584},
  [\href{https://arxiv.org/abs/1902.00134}{{\texttt 1902.00134}}].

\bibitem{Abouabid:2021yvw}
H.~Abouabid, A.~Arhrib, D.~Azevedo, J.~E. Falaki, P.~M. Ferreira,
  M.~M\"uhlleitner et~al., \emph{{Benchmarking di-Higgs production in various
  extended Higgs sector models}},
  \href{http://dx.doi.org/10.1007/JHEP09(2022)011}{\emph{JHEP} \textbf{ 09}
  (2022) 011}, [\href{https://arxiv.org/abs/2112.12515}{{\texttt 2112.12515}}].

\bibitem{Lee:1973iz}
T.~D. Lee, \emph{{A Theory of Spontaneous T Violation}},
  \href{http://dx.doi.org/10.1103/PhysRevD.8.1226}{\emph{Phys. Rev. D} \textbf{
  8} (1973) 1226--1239}.

\bibitem{Branco:2011iw}
G.~C. Branco, P.~M. Ferreira, L.~Lavoura, M.~N. Rebelo, M.~Sher and J.~P.
  Silva, \emph{{Theory and phenomenology of two-Higgs-doublet models}},
  \href{http://dx.doi.org/10.1016/j.physrep.2012.02.002}{\emph{Phys. Rept.}
  \textbf{ 516} (2012) 1--102},
  [\href{https://arxiv.org/abs/1106.0034}{{\texttt 1106.0034}}].

\bibitem{Crivellin:2016ihg}
A.~Crivellin, M.~Ghezzi and M.~Procura, \emph{{Effective Field Theory with Two
  Higgs Doublets}},
  \href{http://dx.doi.org/10.1007/JHEP09(2016)160}{\emph{JHEP} \textbf{ 09}
  (2016) 160}, [\href{https://arxiv.org/abs/1608.00975}{{\texttt 1608.00975}}].

\bibitem{Gunion:1989we}
J.~F. Gunion, H.~E. Haber, G.~L. Kane and S.~Dawson, \emph{{The Higgs Hunter's
  Guide}}, vol.~80.
\newblock Front. Phys., 2000.

\bibitem{Martin:1997ns}
S.~P. Martin, \emph{{A Supersymmetry primer}},
  \href{http://dx.doi.org/10.1142/9789812839657_0001}{\emph{Adv. Ser. Direct.
  High Energy Phys.} \textbf{ 18} (1998) 1--98},
  [\href{https://arxiv.org/abs/hep-ph/9709356}{{\texttt hep-ph/9709356}}].

\bibitem{Dawson:1997tz}
S.~Dawson, \emph{{The MSSM and why it works}},  in \emph{{Theoretical Advanced
  Study Institute in Elementary Particle Physics (TASI 97): Supersymmetry,
  Supergravity and Supercolliders}}, pp.~261--339, 6, 1997.
\newblock \href{https://arxiv.org/abs/hep-ph/9712464}{{\texttt
  hep-ph/9712464}}.

\bibitem{Djouadi:2005gj}
A.~Djouadi, \emph{{The Anatomy of electro-weak symmetry breaking. II. The Higgs
  bosons in the minimal supersymmetric model}},
  \href{http://dx.doi.org/10.1016/j.physrep.2007.10.005}{\emph{Phys. Rept.}
  \textbf{ 459} (2008) 1--241},
  [\href{https://arxiv.org/abs/hep-ph/0503173}{{\texttt hep-ph/0503173}}].

\bibitem{Moretti:2019ulc}
S.~Moretti and S.~Khalil, \emph{{Supersymmetry Beyond Minimality: From Theory
  to Experiment}}.
\newblock CRC Press, 2019.

\bibitem{DeCurtis:2018zvh}
S.~De~Curtis, L.~Delle~Rose, S.~Moretti and K.~Yagyu, \emph{{A Concrete
  Composite 2-Higgs Doublet Model}},
  \href{http://dx.doi.org/10.1007/JHEP12(2018)051}{\emph{JHEP} \textbf{ 12}
  (2018) 051}, [\href{https://arxiv.org/abs/1810.06465}{{\texttt 1810.06465}}].

\bibitem{DeCurtis:2018iqd}
S.~De~Curtis, L.~Delle~Rose, S.~Moretti and K.~Yagyu, \emph{{Supersymmetry
  versus Compositeness: 2HDMs tell the story}},
  \href{http://dx.doi.org/10.1016/j.physletb.2018.09.042}{\emph{Phys. Lett. B}
  \textbf{ 786} (2018) 189--194},
  [\href{https://arxiv.org/abs/1803.01865}{{\texttt 1803.01865}}].

\bibitem{Moretti:2023dlx}
S.~Moretti, L.~Panizzi, J.~Sj\"olin and H.~Waltari, \emph{{Deconstructing
  squark contributions to di-Higgs production at the LHC}},
  \href{http://dx.doi.org/10.1103/PhysRevD.107.115010}{\emph{Phys. Rev. D}
  \textbf{ 107} (2023) 115010},
  [\href{https://arxiv.org/abs/2302.03401}{{\texttt 2302.03401}}].

\bibitem{Gabriel:2023dyx}
P.~Gabriel, M.~M\"uhlleitner, D.~Neacsu and R.~Santos, \emph{{Dark Coloured
  Scalars Impact on Single and Di-Higgs Production at the LHC}},
  \href{https://arxiv.org/abs/2308.07023}{{\texttt 2308.07023}}.

\bibitem{Cheung_2021}
K.~Cheung, A.~Jueid, C.-T. Lu, J.~Song and Y.~W. Yoon, \emph{Disentangling new
  physics effects on nonresonant higgs boson pair production from gluon
  fusion}, \href{http://dx.doi.org/10.1103/physrevd.103.015019}{\emph{Physical
  Review D} \textbf{ 103} (jan, 2021) }.

\bibitem{Contino:2010rs}
R.~Contino, \emph{{The Higgs as a Composite Nambu-Goldstone Boson}},  in
  \emph{{Theoretical Advanced Study Institute in Elementary Particle Physics}:
  {Physics of the Large and the Small}}, pp.~235--306, 2011.
\newblock \href{https://arxiv.org/abs/1005.4269}{{\texttt 1005.4269}}.
\newblock \href{http://dx.doi.org/10.1142/9789814327183_0005}{DOI}.

\bibitem{Dimopoulos:1981xc}
S.~Dimopoulos and J.~Preskill, \emph{{Massless Composites With Massive
  Constituents}},
  \href{http://dx.doi.org/10.1016/0550-3213(82)90345-5}{\emph{Nucl. Phys. B}
  \textbf{ 199} (1982) 206--222}.

\bibitem{Kaplan:1983sm}
D.~B. Kaplan, H.~Georgi and S.~Dimopoulos, \emph{{Composite Higgs Scalars}},
  \href{http://dx.doi.org/10.1016/0370-2693(84)91178-X}{\emph{Phys. Lett. B}
  \textbf{ 136} (1984) 187--190}.

\bibitem{Kaplan:1983fs}
D.~B. Kaplan and H.~Georgi, \emph{{SU(2) x U(1) Breaking by Vacuum
  Misalignment}},
  \href{http://dx.doi.org/10.1016/0370-2693(84)91177-8}{\emph{Phys. Lett. B}
  \textbf{ 136} (1984) 183--186}.

\bibitem{Banks:1984gj}
T.~Banks, \emph{{Constraints on SU2×U1 breaking by vacuum misalignment}},
  \href{http://dx.doi.org/10.1016/0550-3213(84)90389-4}{\emph{Nucl. Phys. B}
  \textbf{ 243} (1984) 125--130}.

\bibitem{Georgi:1984ef}
H.~Georgi, D.~B. Kaplan and P.~Galison, \emph{{Calculation of the Composite
  Higgs Mass}},
  \href{http://dx.doi.org/10.1016/0370-2693(84)90823-2}{\emph{Phys. Lett. B}
  \textbf{ 143} (1984) 152--154}.

\bibitem{Georgi:1984af}
H.~Georgi and D.~B. Kaplan, \emph{{Composite Higgs and Custodial SU(2)}},
  \href{http://dx.doi.org/10.1016/0370-2693(84)90341-1}{\emph{Phys. Lett. B}
  \textbf{ 145} (1984) 216--220}.

\bibitem{Dugan:1984hq}
M.~J. Dugan, H.~Georgi and D.~B. Kaplan, \emph{{Anatomy of a Composite Higgs
  Model}}, \href{http://dx.doi.org/10.1016/0550-3213(85)90221-4}{\emph{Nucl.
  Phys. B} \textbf{ 254} (1985) 299--326}.

\bibitem{Mrazek:2011iu}
J.~Mrazek, A.~Pomarol, R.~Rattazzi, M.~Redi, J.~Serra and A.~Wulzer, \emph{{The
  Other Natural Two Higgs Doublet Model}},
  \href{http://dx.doi.org/10.1016/j.nuclphysb.2011.07.008}{\emph{Nucl. Phys. B}
  \textbf{ 853} (2011) 1--48}, [\href{https://arxiv.org/abs/1105.5403}{{\texttt
  1105.5403}}].

\bibitem{DeCurtis:2016scv}
S.~De~Curtis, S.~Moretti, K.~Yagyu and E.~Yildirim, \emph{{Perturbative
  unitarity bounds in composite two-Higgs doublet models}},
  \href{http://dx.doi.org/10.1103/PhysRevD.94.055017}{\emph{Phys. Rev. D}
  \textbf{ 94} (2016) 055017},
  [\href{https://arxiv.org/abs/1602.06437}{{\texttt 1602.06437}}].

\bibitem{DeCurtis:2016tsm}
S.~De~Curtis, S.~Moretti, K.~Yagyu and E.~Yildirim, \emph{{LHC Phenomenology of
  Composite 2-Higgs Doublet Models}},
  \href{http://dx.doi.org/10.1140/epjc/s10052-017-5082-4}{\emph{Eur. Phys. J.
  C} \textbf{ 77} (2017) 513},
  [\href{https://arxiv.org/abs/1610.02687}{{\texttt 1610.02687}}].

\bibitem{DeCurtis:2017gzi}
S.~De~Curtis, S.~Moretti, K.~Yagyu and E.~Yildirim, \emph{{Single and double
  SM-like Higgs boson production at future electron-positron colliders in
  composite 2HDMs}},
  \href{http://dx.doi.org/10.1103/PhysRevD.95.095026}{\emph{Phys. Rev. D}
  \textbf{ 95} (2017) 095026},
  [\href{https://arxiv.org/abs/1702.07260}{{\texttt 1702.07260}}].

\bibitem{Kaplan:1991dc}
D.~B. Kaplan, \emph{{Flavor at SSC energies: A New mechanism for dynamically
  generated fermion masses}},
  \href{http://dx.doi.org/10.1016/S0550-3213(05)80021-5}{\emph{Nucl. Phys. B}
  \textbf{ 365} (1991) 259--278}.

\bibitem{DeCurtis:2011yx}
S.~De~Curtis, M.~Redi and A.~Tesi, \emph{{The 4D Composite Higgs}},
  \href{http://dx.doi.org/10.1007/JHEP04(2012)042}{\emph{JHEP} \textbf{ 04}
  (2012) 042}, [\href{https://arxiv.org/abs/1110.1613}{{\texttt 1110.1613}}].

\bibitem{DeCurtis:2021uqx}
S.~De~Curtis, S.~Moretti, R.~Nagai and K.~Yagyu, \emph{{CP-Violation in a
  composite 2-Higgs doublet model}},
  \href{http://dx.doi.org/10.1007/JHEP10(2021)040}{\emph{JHEP} \textbf{ 10}
  (2021) 040}, [\href{https://arxiv.org/abs/2107.08201}{{\texttt 2107.08201}}].

\bibitem{Bechtle:2013wla}
P.~Bechtle, O.~Brein, S.~Heinemeyer, O.~St\r{a}l, T.~Stefaniak, G.~Weiglein
  et~al., \emph{{$\mathsf{HiggsBounds}-4$: Improved Tests of Extended Higgs
  Sectors against Exclusion Bounds from LEP, the Tevatron and the LHC}},
  \href{http://dx.doi.org/10.1140/epjc/s10052-013-2693-2}{\emph{Eur. Phys. J.
  C} \textbf{ 74} (2014) 2693},
  [\href{https://arxiv.org/abs/1311.0055}{{\texttt 1311.0055}}].

\bibitem{Bechtle:2013xfa}
P.~Bechtle, S.~Heinemeyer, O.~St\r{a}l, T.~Stefaniak and G.~Weiglein,
  \emph{{$HiggsSignals$: Confronting arbitrary Higgs sectors with measurements
  at the Tevatron and the LHC}},
  \href{http://dx.doi.org/10.1140/epjc/s10052-013-2711-4}{\emph{Eur. Phys. J.
  C} \textbf{ 74} (2014) 2711},
  [\href{https://arxiv.org/abs/1305.1933}{{\texttt 1305.1933}}].

\bibitem{DeCurtis:2019jwg}
S.~De~Curtis, L.~Delle~Rose, S.~Moretti and K.~Yagyu, \emph{{A Composite
  2-Higgs Doublet Model}},
  \href{http://dx.doi.org/10.22323/1.364.0344}{\emph{PoS} \textbf{ EPS-HEP2019}
  (2020) 344}, [\href{https://arxiv.org/abs/1910.13699}{{\texttt 1910.13699}}].

\bibitem{deBlas:2019rxi}
J.~de~Blas et~al., \emph{{Higgs Boson Studies at Future Particle Colliders}},
  \href{http://dx.doi.org/10.1007/JHEP01(2020)139}{\emph{JHEP} \textbf{ 01}
  (2020) 139}, [\href{https://arxiv.org/abs/1905.03764}{{\texttt 1905.03764}}].

\bibitem{Benbrik:2019zdp}
R.~Benbrik et~al., \emph{{Signatures of vector-like top partners decaying into
  new neutral scalar or pseudoscalar bosons}},
  \href{http://dx.doi.org/10.1007/JHEP05(2020)028}{\emph{JHEP} \textbf{ 05}
  (2020) 028}, [\href{https://arxiv.org/abs/1907.05929}{{\texttt 1907.05929}}].

\bibitem{ATLAS:2022tla}
{\scshape ATLAS} collaboration, \emph{{Search for pair-produced vector-like top
  and bottom partners in events with large missing transverse momentum in pp
  collisions with the ATLAS detector}},
  \href{https://arxiv.org/abs/2212.05263}{{\texttt 2212.05263}}.

\bibitem{Grober:2010yv}
R.~Grober and M.~M{\"u}hlleitner, \emph{{Composite Higgs Boson Pair Production
  at the LHC}}, \href{http://dx.doi.org/10.1007/JHEP06(2011)020}{\emph{JHEP}
  \textbf{ 06} (2011) 020}, [\href{https://arxiv.org/abs/1012.1562}{{\texttt
  1012.1562}}].

\bibitem{Gillioz:2012se}
M.~Gillioz, R.~Grober, C.~Grojean, M.~M{\"u}hlleitner and E.~Salvioni,
  \emph{{Higgs Low-Energy Theorem (and its corrections) in Composite Models}},
  \href{http://dx.doi.org/10.1007/JHEP10(2012)004}{\emph{JHEP} \textbf{ 10}
  (2012) 004}, [\href{https://arxiv.org/abs/1206.7120}{{\texttt 1206.7120}}].

\bibitem{Grober:2015cwa}
R.~Grober, M.~M{\"u}hlleitner, M.~Spira and J.~Streicher, \emph{{NLO QCD
  Corrections to Higgs Pair Production including Dimension-6 Operators}},
  \href{http://dx.doi.org/10.1007/JHEP09(2015)092}{\emph{JHEP} \textbf{ 09}
  (2015) 092}, [\href{https://arxiv.org/abs/1504.06577}{{\texttt 1504.06577}}].

\bibitem{Grober:2016wmf}
R.~Grober, M.~M{\"u}hlleitner and M.~Spira, \emph{{Signs of Composite Higgs
  Pair Production at Next-to-Leading Order}},
  \href{http://dx.doi.org/10.1007/JHEP06(2016)080}{\emph{JHEP} \textbf{ 06}
  (2016) 080}, [\href{https://arxiv.org/abs/1602.05851}{{\texttt 1602.05851}}].

\bibitem{HPair}
M.~Spira, ``{HPair}.'' \url{http://tiger.web.psi.ch/proglist.html}.

\bibitem{vanOldenborgh:1989wn}
G.~J. van Oldenborgh and J.~A.~M. Vermaseren, \emph{{New Algorithms for One
  Loop Integrals}}, \href{http://dx.doi.org/10.1007/BF01621031}{\emph{Z. Phys.
  C} \textbf{ 46} (1990) 425--438}.

\bibitem{Hahn:1998yk}
T.~Hahn and M.~Perez-Victoria, \emph{{Automatized one loop calculations in
  four-dimensions and D-dimensions}},
  \href{http://dx.doi.org/10.1016/S0010-4655(98)00173-8}{\emph{Comput. Phys.
  Commun.} \textbf{ 118} (1999) 153--165},
  [\href{https://arxiv.org/abs/hep-ph/9807565}{{\texttt hep-ph/9807565}}].

\bibitem{Carrazza:2016gav}
S.~Carrazza, R.~K. Ellis and G.~Zanderighi, \emph{{QCDLoop: a comprehensive
  framework for one-loop scalar integrals}},
  \href{http://dx.doi.org/10.1016/j.cpc.2016.07.033}{\emph{Comput. Phys.
  Commun.} \textbf{ 209} (2016) 134--143},
  [\href{https://arxiv.org/abs/1605.03181}{{\texttt 1605.03181}}].

\bibitem{Djouadi:1997yw}
A.~Djouadi, J.~Kalinowski and M.~Spira, \emph{{HDECAY: A Program for Higgs
  boson decays in the standard model and its supersymmetric extension}},
  \href{http://dx.doi.org/10.1016/S0010-4655(97)00123-9}{\emph{Comput. Phys.
  Commun.} \textbf{ 108} (1998) 56--74},
  [\href{https://arxiv.org/abs/hep-ph/9704448}{{\texttt hep-ph/9704448}}].

\bibitem{Djouadi:2018xqq}
A.~Djouadi, J.~Kalinowski, M.~M{\"u}hlleitner and M.~Spira, \emph{{HDECAY:
  Twenty$_{++}$ years after}},
  \href{http://dx.doi.org/10.1016/j.cpc.2018.12.010}{\emph{Comput. Phys.
  Commun.} \textbf{ 238} (2019) 214--231},
  [\href{https://arxiv.org/abs/1801.09506}{{\texttt 1801.09506}}].

\bibitem{Spira_1995}
M.~Spira, A.~Djouadi, D.~Graudenz and R.~Zerwas, \emph{Higgs boson production
  at the {LHC}},
  \href{http://dx.doi.org/10.1016/0550-3213(95)00379-7}{\emph{Nuclear Physics
  B} \textbf{ 453} (oct, 1995) 17--82}.

\bibitem{spira1995higlu}
M.~Spira, \emph{{HIGLU}: A program for the calculation of the total higgs
  production cross section at hadron colliders via gluon fusion including qcd
  corrections},  1995.

\bibitem{LHCHiggsCrossSectionWorkingGroup:2011wcg}
{\scshape LHC Higgs Cross Section Working Group} collaboration, S.~Dittmaier
  et~al., \emph{{Handbook of LHC Higgs Cross Sections: 1. Inclusive
  Observables}},  \href{https://arxiv.org/abs/1101.0593}{{\texttt 1101.0593}}.

\bibitem{Dittmaier:2012vm}
S.~Dittmaier et~al., \emph{{Handbook of LHC Higgs Cross Sections: 2.
  Differential Distributions}},
  \href{https://arxiv.org/abs/1201.3084}{{\texttt 1201.3084}}.

\bibitem{LHCHiggsCrossSectionWorkingGroup:2016ypw}
{\scshape LHC Higgs Cross Section Working Group} collaboration, D.~de~Florian
  et~al., \emph{{Handbook of LHC Higgs Cross Sections: 4. Deciphering the
  Nature of the Higgs Sector}},
  \href{https://arxiv.org/abs/1610.07922}{{\texttt 1610.07922}}.

\bibitem{Spira:2016ztx}
M.~Spira, \emph{{Higgs Boson Production and Decay at Hadron Colliders}},
  \href{http://dx.doi.org/10.1016/j.ppnp.2017.04.001}{\emph{Prog. Part. Nucl.
  Phys.} \textbf{ 95} (2017) 98--159},
  [\href{https://arxiv.org/abs/1612.07651}{{\texttt 1612.07651}}].

\bibitem{Heinrich:2020ybq}
G.~Heinrich, \emph{{Collider Physics at the Precision Frontier}},
  \href{http://dx.doi.org/10.1016/j.physrep.2021.03.006}{\emph{Phys. Rept.}
  \textbf{ 922} (2021) 1--69},
  [\href{https://arxiv.org/abs/2009.00516}{{\texttt 2009.00516}}].

\bibitem{Nhung:2013lpa}
D.~T. Nhung, M.~M{\"u}hlleitner, J.~Streicher and K.~Walz, \emph{{Higher Order
  Corrections to the Trilinear Higgs Self-Couplings in the Real NMSSM}},
  \href{http://dx.doi.org/10.1007/JHEP11(2013)181}{\emph{JHEP} \textbf{ 11}
  (2013) 181}, [\href{https://arxiv.org/abs/1306.3926}{{\texttt 1306.3926}}].

\bibitem{Grober:2017gut}
R.~Grober, M.~M{\"u}hlleitner and M.~Spira, \emph{{Higgs Pair Production at NLO
  QCD for CP-violating Higgs Sectors}},
  \href{http://dx.doi.org/10.1016/j.nuclphysb.2017.10.002}{\emph{Nucl. Phys. B}
  \textbf{ 925} (2017) 1--27},
  [\href{https://arxiv.org/abs/1705.05314}{{\texttt 1705.05314}}].

\bibitem{Borschensky:2022pfc}
C.~Borschensky, T.~N. Dao, M.~Gabelmann, M.~M\"uhlleitner and H.~Rzehak,
  \emph{{The trilinear Higgs self-couplings at $\mathcal {O}(\alpha _t^2)$ in
  the CP-violating NMSSM}},
  \href{http://dx.doi.org/10.1140/epjc/s10052-023-11215-5}{\emph{Eur. Phys. J.
  C} \textbf{ 83} (2023) 118},
  [\href{https://arxiv.org/abs/2210.02104}{{\texttt 2210.02104}}].

\bibitem{Arco:2022lai}
F.~Arco, S.~Heinemeyer, M.~M\"uhlleitner and K.~Radchenko, \emph{{Sensitivity
  to Triple Higgs Couplings via Di-Higgs Production in the 2HDM at the
  (HL-)LHC}},  \href{https://arxiv.org/abs/2212.11242}{{\texttt 2212.11242}}.

\bibitem{Baglio:2023euv}
J.~Baglio, F.~Campanario, S.~Glaus, M.~M\"uhlleitner, J.~Ronca and M.~Spira,
  \emph{{Full NLO QCD predictions for Higgs-pair production in the
  2-Higgs-Doublet Model}},  \href{https://arxiv.org/abs/2303.05409}{{\texttt
  2303.05409}}.

\bibitem{Harland_Lang_2015}
L.~A. Harland-Lang, A.~D. Martin, P.~Motylinski and R.~S. Thorne, \emph{Parton
  distributions in the {LHC} era: {MMHT} 2014 {PDFs}},
  \href{http://dx.doi.org/10.1140/epjc/s10052-015-3397-6}{\emph{The European
  Physical Journal C} \textbf{ 75} (may, 2015) }.

\bibitem{CMS:2021qvd}
{\scshape CMS} collaboration, \emph{{Search for resonant Higgs boson pair
  production in four b quark final state using large-area jets in proton-proton
  collisions at $\sqrt{s}=13~\mathrm{TeV}$}}, {\emph{CMS-PAS-B2G-20-004} (2021)
  }.

\bibitem{ATLAS:2021ifb}
{\scshape ATLAS} collaboration, G.~Aad et~al., \emph{{Search for Higgs boson
  pair production in the two bottom quarks plus two photons final state in $pp$
  collisions at $\sqrt{s}=13$ TeV with the ATLAS detector}},
  \href{http://dx.doi.org/10.1103/PhysRevD.106.052001}{\emph{Phys. Rev. D}
  \textbf{ 106} (2022) 052001},
  [\href{https://arxiv.org/abs/2112.11876}{{\texttt 2112.11876}}].

\bibitem{ATLAS:2022hwc}
{\scshape ATLAS} collaboration, G.~Aad et~al., \emph{{Search for resonant pair
  production of Higgs bosons in the $b\bar{b}b\bar{b}$ final state using $pp$
  collisions at $\sqrt{s}$ = 13 TeV with the ATLAS detector}},
  \href{http://dx.doi.org/10.1103/PhysRevD.105.092002}{\emph{Phys. Rev. D}
  \textbf{ 105} (2022) 092002},
  [\href{https://arxiv.org/abs/2202.07288}{{\texttt 2202.07288}}].

\bibitem{ATLAS:2022xzm}
{\scshape ATLAS} collaboration, G.~Aad et~al., \emph{{Search for resonant and
  non-resonant Higgs boson pair production in the $
  b\overline{b}{\tau}^{+}{\tau}^{-} $ decay channel using 13 TeV pp collision
  data from the ATLAS detector}},
  \href{http://dx.doi.org/10.1007/JHEP07(2023)040}{\emph{JHEP} \textbf{ 07}
  (2023) 040}, [\href{https://arxiv.org/abs/2209.10910}{{\texttt 2209.10910}}].

\bibitem{ATLAS:2018rnh}
M.~A.~{\em et~al.}. ATLAS, \emph{Search for pair production of higgs bosons in
  the $b\overline{b}b\overline{b}$ final state using proton-proton collisions
  at $ \sqrt{s}=13 $ {TeV} with the {ATLAS} detector},
  \href{http://dx.doi.org/10.1007/jhep01(2019)030}{\emph{Journal of High Energy
  Physics} \textbf{ 2019} (Jan, 2019) }.

\bibitem{ATLAS:2018uni}
M.~A.~{\em et~al.}. ATLAS, \emph{Search for resonant and nonresonant higgs
  boson pair production in the
  $b\overline{b}{\ensuremath{\tau}}^{+}{\ensuremath{\tau}}^{\ensuremath{-}}$
  decay channel in $pp$ collisions at $\sqrt{s}=13\text{ }\text{ }\mathrm{TeV}$
  with the {ATLAS} detector},
  \href{http://dx.doi.org/10.1103/physrevlett.121.191801}{\emph{Physical Review
  Letters} \textbf{ 121} (Nov, 2018) }.

\bibitem{PhysRevLett.122.089901}
{\scshape ATLAS Collaboration} collaboration, M.~A.~{\em et~al.}. ATLAS,
  \emph{Erratum: Search for resonant and nonresonant higgs boson pair
  production in the
  $b\overline{b}{\ensuremath{\tau}}^{+}{\ensuremath{\tau}}^{\ensuremath{-}}$
  decay channel in $pp$ collisions at $\sqrt{s}=13\text{ }\text{ }\mathrm{TeV}$
  with the {ATLAS} detector [phys. rev. lett. 121, 191801 (2018)]},
  \href{http://dx.doi.org/10.1103/PhysRevLett.122.089901}{\emph{Phys. Rev.
  Lett.} \textbf{ 122} (Feb, 2019) 089901}.

\bibitem{ATLAS:2020azv}
G.~A.~{\em et~al.}. ATLAS, \emph{Reconstruction and identification of boosted
  di-$\tau$ systems in a search for higgs boson pairs using 13 tev
  proton-proton collision data in {ATLAS}},
  \href{http://dx.doi.org/10.1007/JHEP11(2020)163}{\emph{Journal of High Energy
  Physics} \textbf{ 2020} (Nov, 2020) 163}.

\bibitem{ATLAS:2018dpp}
M.~A.~{\em et~al.}. ATLAS, \emph{Search for higgs boson pair production in the
  $ \gamma \gamma b\overline{b}$ final state with 13 {TeV} pp collision data
  collected by the {ATLAS} experiment},
  \href{http://dx.doi.org/10.1007/jhep11(2018)040}{\emph{Journal of High Energy
  Physics} \textbf{ 2018} (Nov, 2018) }.

\bibitem{ATLAS:2018fpd}
M.~A.~{\em et~al.}. ATLAS, \emph{Search for higgs boson pair production in the
  $b\overline{b} {W} {W}^{\ast} $ decay mode at $ \sqrt{s} $ = 13 {TeV} with
  the {ATLAS} detector},
  \href{http://dx.doi.org/10.1007/jhep04(2019)092}{\emph{Journal of High Energy
  Physics} \textbf{ 2019} (Apr, 2019) }.

\bibitem{CMS:2020jeo}
A.~M. S.~{\em et~al.}. CMS, \emph{Search for resonant pair production of higgs
  bosons in the $bb{ZZ}$ channel in proton-proton collisions at
  $\sqrt{s}=13\text{ }\text{}\mathrm{TeV}$},
  \href{http://dx.doi.org/10.1103/physrevd.102.032003}{\emph{Physical Review D}
  \textbf{ 102} (Aug, 2020) }.

\bibitem{ATLAS:2018hqk}
M.~A.~{\em et~al.}. ATLAS, \emph{Search for higgs boson pair production in the
  $\gamma \gamma {WW}^{\ast}$ channel using $pp$ collision data recorded at $
  \sqrt{s}=13$ {TeV} with the {ATLAS} detector},
  \href{http://dx.doi.org/10.1140/epjc/s10052-018-6457-x}{\emph{The European
  Physical Journal C} \textbf{ 78} (Dec, 2018) }.

\bibitem{ATLAS:2018ili}
M.~A.~{\em et~al.}. ATLAS, \emph{Search for higgs boson pair production in the
  ${WW}^{({\ast})}{WW}^{({\ast})}$ decay channel using {ATLAS} data recorded at
  $ \sqrt{\mathrm{s}}$ = 13 {TeV}},
  \href{http://dx.doi.org/10.1007/jhep05(2019)124}{\emph{Journal of High Energy
  Physics} \textbf{ 2019} (May, 2019) }.

\bibitem{ATLAS:2022jtk}
{\scshape ATLAS} collaboration, G.~Aad et~al., \emph{{Constraints on the Higgs
  boson self-coupling from single- and double-Higgs production with the ATLAS
  detector using pp collisions at $\sqrt{s}=$13 TeV}},
  \href{http://dx.doi.org/10.1016/j.physletb.2023.137745}{\emph{Phys. Lett. B}
  \textbf{ 843} (2023) 137745},
  [\href{https://arxiv.org/abs/2211.01216}{{\texttt 2211.01216}}].

\bibitem{Pich:2009sp}
A.~Pich and P.~Tuzon, \emph{{Yukawa Alignment in the Two-Higgs-Doublet Model}},
  \href{http://dx.doi.org/10.1103/PhysRevD.80.091702}{\emph{Phys. Rev. D}
  \textbf{ 80} (2009) 091702}, [\href{https://arxiv.org/abs/0908.1554}{{\texttt
  0908.1554}}].

\bibitem{Staub:2008uz}
F.~Staub, \emph{{SARAH}},  \href{https://arxiv.org/abs/0806.0538}{{\texttt
  0806.0538}}.

\bibitem{Davidson:2005cw}
S.~Davidson and H.~E. Haber, \emph{{Basis-independent methods for the
  two-Higgs-doublet model}},
  \href{http://dx.doi.org/10.1103/PhysRevD.72.099902}{\emph{Phys. Rev. D}
  \textbf{ 72} (2005) 035004},
  [\href{https://arxiv.org/abs/hep-ph/0504050}{{\texttt hep-ph/0504050}}].

\bibitem{Haber:1999zh}
H.~E. Haber and H.~E. Logan, \emph{{Radiative corrections to the Z b anti-b
  vertex and constraints on extended Higgs sectors}},
  \href{http://dx.doi.org/10.1103/PhysRevD.62.015011}{\emph{Phys. Rev. D}
  \textbf{ 62} (2000) 015011},
  [\href{https://arxiv.org/abs/hep-ph/9909335}{{\texttt hep-ph/9909335}}].

\bibitem{Deschamps:2009rh}
O.~Deschamps, S.~Descotes-Genon, S.~Monteil, V.~Niess, S.~T'Jampens and
  V.~Tisserand, \emph{{The Two Higgs Doublet of Type II facing flavour physics
  data}}, \href{http://dx.doi.org/10.1103/PhysRevD.82.073012}{\emph{Phys. Rev.
  D} \textbf{ 82} (2010) 073012},
  [\href{https://arxiv.org/abs/0907.5135}{{\texttt 0907.5135}}].

\bibitem{Mahmoudi:2009zx}
F.~Mahmoudi and O.~Stal, \emph{{Flavor constraints on the two-Higgs-doublet
  model with general Yukawa couplings}},
  \href{http://dx.doi.org/10.1103/PhysRevD.81.035016}{\emph{Phys. Rev. D}
  \textbf{ 81} (2010) 035016}, [\href{https://arxiv.org/abs/0907.1791}{{\texttt
  0907.1791}}].

\bibitem{Hermann:2012fc}
T.~Hermann, M.~Misiak and M.~Steinhauser, \emph{{$\bar{B}\to X_s \gamma$ in the
  Two Higgs Doublet Model up to Next-to-Next-to-Leading Order in QCD}},
  \href{http://dx.doi.org/10.1007/JHEP11(2012)036}{\emph{JHEP} \textbf{ 11}
  (2012) 036}, [\href{https://arxiv.org/abs/1208.2788}{{\texttt 1208.2788}}].

\bibitem{Misiak:2015xwa}
M.~Misiak et~al., \emph{{Updated NNLO QCD predictions for the weak radiative
  B-meson decays}},
  \href{http://dx.doi.org/10.1103/PhysRevLett.114.221801}{\emph{Phys. Rev.
  Lett.} \textbf{ 114} (2015) 221801},
  [\href{https://arxiv.org/abs/1503.01789}{{\texttt 1503.01789}}].

\bibitem{Misiak:2017bgg}
M.~Misiak and M.~Steinhauser, \emph{{Weak radiative decays of the B meson and
  bounds on $M_{H^\pm }$ in the Two-Higgs-Doublet Model}},
  \href{http://dx.doi.org/10.1140/epjc/s10052-017-4776-y}{\emph{Eur. Phys. J.
  C} \textbf{ 77} (2017) 201},
  [\href{https://arxiv.org/abs/1702.04571}{{\texttt 1702.04571}}].

\bibitem{Misiak:2020vlo}
M.~Misiak, A.~Rehman and M.~Steinhauser, \emph{{Towards $ \overline{B}\to
  {X}_s\gamma $ at the NNLO in QCD without interpolation in m$_{c}$}},
  \href{http://dx.doi.org/10.1007/JHEP06(2020)175}{\emph{JHEP} \textbf{ 06}
  (2020) 175}, [\href{https://arxiv.org/abs/2002.01548}{{\texttt 2002.01548}}].

\bibitem{DeCurtis:2016gly}
S.~De~Curtis, S.~Moretti, K.~Yagyu and E.~Yildirim, \emph{{Theory and
  Phenomenology of Composite 2-Higgs Doublet Models}},
  \href{http://dx.doi.org/10.22323/1.286.0018}{\emph{PoS} \textbf{ CHARGED2016}
  (2016) 018}, [\href{https://arxiv.org/abs/1612.05125}{{\texttt 1612.05125}}].

\end{thebibliography}\endgroup
\bibliographystyle{JHEP}

\end{document}